\documentclass[article,twocolumn,superscriptaddress,bibnotes,nofootinbib]{revtex4-2}

\usepackage[normalem]{ulem}

\usepackage{mathtools}
\usepackage{array}
\usepackage{hyperref}
\usepackage{amsmath,amssymb}
\usepackage{longtable}

\DeclareMathOperator{\dd}{d\hspace{-2pt}}
\DeclareMathOperator{\dx}{d\textbf{x}}
\DeclareMathOperator{\Pe}{Pe}

\begin{document}

\title{\textbf{Modes of Mechanical Guidance of Adhesion-Independent Cell Migration} }

\author{Hanna Luise Gertack}
\affiliation{ Institute of Numerical Mathematics and Optimization, TU Bergakademie Freiberg, Germany}%

\author{Peter A.E. Hampshire}
\affiliation{Max Planck Institute for the Physics of Complex Systems, Dresden, Germany}
\affiliation{Center for Systems Biology Dresden, Dresden, Germany}

\author{Claudia Wohlgemuth}
\affiliation{ Institute of Numerical Mathematics and Optimization, TU Bergakademie Freiberg, Germany}%

\author{Ricard Alert}
\affiliation{Max Planck Institute for the Physics of Complex Systems, Dresden, Germany}
\affiliation{Center for Systems Biology Dresden, Dresden, Germany}
\affiliation{Cluster of Excellence Physics of Life, TU Dresden, Germany}
\affiliation{Departament de Física de la Matèria Condensada, Universitat de Barcelona, Barcelona, Spain}
\affiliation{Universitat de Barcelona Institute of Complex Systems (UBICS), Barcelona, Spain}
\affiliation{Institució Catalana de Recerca i Estudis Avançats (ICREA), Barcelona, Spain}

\author{Sebastian Aland* }
\affiliation{ Institute of Numerical Mathematics and Optimization, TU Bergakademie Freiberg, Germany}%
\affiliation{Center for Systems Biology Dresden, Dresden, Germany}
\affiliation{Faculty of Computer Science/Mathematics, HTW Dresden, Germany}

\begin{abstract}
Adhesion-independent migration is a prominent mode of cell motility in confined environments, yet the physical principles that guide such movement remain incompletely understood. We present a phase-field model for simulating the motility of deformable, non-adherent cells driven by contractile surface instabilities of the cell cortex. This model couples surface and bulk hydrodynamics, accommodates large shape deformations and incorporates a diffusible contraction-generating molecule (myosin) that drives cortical flows. These capabilities enable a systematic exploration of how mechanical cues direct cell polarization and migration.
We first demonstrate that spontaneous symmetry breaking of cortical activity can lead to persistent and directed movement in channels. We then investigate how various physical cues - including gradients in friction, viscosity, and channel width as well as external flows and hydrodynamic interactions between cells - steer migration. Our results show that active surface dynamics can generate stimulus-specific cell behaviors, such as migration up friction gradients or escape from narrow regions.
Beyond cell migration, the model offers a versatile platform for exploring the mechanics of active surfaces in biological systems.
\end{abstract}

\footnotetext{Corresponding author at: Institute of Numerical Mathematics and Optimization, TU Bergakademie Freiberg, Akademiestr. 6, 09599 Freiberg, Germany. 
\href{mailto:sebastian.aland@math.tu-freiberg.de}{sebastian.aland@math.tu-freiberg.de}}

\maketitle

\section{Introduction}

Adhesion-independent migration, commonly referred to as amoeboid migration, plays a key role in development, immune surveillance, and cancer invasion \cite{Bodor.2020,Paluch.2016,Bergert.2015, Yamada.2019,Tozluoglu.2013, Liu.2015, Ruprecht.2015}.  
In vitro studies have shown that this migratory mode can be induced in a wide range of cell types when subjected to three-dimensional confinement \cite{Liu.2015, Ruprecht.2015,Kang.2024}.
Although there are alternative mechanisms \cite{ Stroka.2014}, amoeboid motility is often driven by actomyosin accumulation at the cell rear, which produces retrograde flows of the cell cortex. This cortical flow then propels the cell through friction with the environment \cite{Bergert.2015,Tozluoglu.2013,Liu.2015,Ruprecht.2015,Singha.2025}.
Although several theoretical frameworks have been proposed to explain how this migration is initiated and sustained, the exact processes governing the symmetry breaking that triggers polarization remain incompletely understood  \cite{Bodor.2020,Paluch.2016,Kang.2024,Liu.2015,Moreau.2018,Singha.2025}. 

Minimal models  \cite{Paluch.2016,Liu.2015,Ruprecht.2015,Bergert.2015,Reversat.2020,Recho.2013, CallanJones.2013,CallanJones.2016,CallanJones.2022,Mietke_PNAS_2019, Mietke_PRL_2019, AlandWittwer_2023,Singha.2025} have suggested that the cell cortex can be described as an active fluid surface that self-organizes tension-generating molecules - such as myosin - into a localized concentration peak through the interplay of advective transport, contractility, and surface flows. If myosin accumulates towards one side of the cell, this side becomes the cell rear and it initiates retrograde flows that propel the cell forward, consistent with amoeboid migration dynamics  \cite{lammermann2009mechanical}. While most prior studies have focused on such self-organization on fixed geometries \cite{Mietke_PRL_2019,Kumar2014-gw,Moore2014-dm,Sehring2015-hg,Weber2018-nn}, experiments and theory have shown that shape changes can actively feed back on cortical flows \cite{Ruprecht.2015,CallanJones.2016_shape}, underscoring the need to incorporate deformable surfaces into modeling frameworks.

Deforming active surfaces were considered in Refs.  \cite{Mietke_PNAS_2019,AlandWittwer_2023, torres2019modelling, Bonati_2022}, coupled to hydrodynamics and material flow. Significant shape deformations including strong constrictions could be reproduced for tubular surfaces  \cite{Mietke_PNAS_2019} as well as for ellipsoidal and spherical surfaces  \cite{AlandWittwer_2023,Bonati_2022}.
However, all previous methods operate with a grid-based representation of the surface. Correspondingly, the surrounding medium was either neglected  \cite{Mietke_PNAS_2019, torres2019modelling} or limited to a simple homogeneous fluid without walls or obstacles  \cite{AlandWittwer_2023,Bonati_2022}. 
Recent work has addressed some of these limitations by modeling the effects of viscous and viscoelastic resistance in confined migration scenarios, providing new predictions on polarization thresholds \cite{Singha.2025}.
In general, the grid-based Lagrangian surface representation of these approaches makes it difficult to include (wall) contact as the mesh typically deforms and tangles in the contact zone and additional contact detection and repulsion algorithms are required. 
Also, grid-based methods are not suited for topological transitions like cell division, which is often modeled as a self-organized process on an active surface. \cite{Mietke_PRL_2019, Bonati_2022}. 

To overcome these limitations, we propose a phase-field approach to represent the evolving surface. The phase-field description makes the model well-suited for simulating migration through complex environments. The model is not only flexible to deal with complex surrounding geometries, including contact to walls and obstacles, but also enables the simulation of topological cell-shape changes, as encountered during self-organized cell division. 
We then use this model to explore how polarization and migration of cells are triggered by various mechanical cues, such as gradients in friction or viscosity, surrounding fluid flow, and constrictions. 
Altogether, we establish a versatile modeling platform for studying active surface dynamics and use it to analyze how cells choose direction in diverse microenvironments.

\section{Active Surface Model}
To develop a diffuse-interface model, we first introduce a sharp-interface formulation describing the dynamics of an active gel surface immersed in a fluid medium.
To this end, we recast the governing equations
given in Refs. \cite{Mietke_PNAS_2019,Mietke_PRL_2019,AlandWittwer_2023} in a form suitable for a later conversion into a diffuse-interface representation. 
Afterward, these equations are non-dimensionalized and the diffuse-interface formulation is introduced, which finally allows us to handle wall contact such as to simulate pattern formation and cell migration in a channel. 

\subsection{Sharp-interface Model}
\begin{figure}
	\centering
	\includegraphics[width=8.5cm]{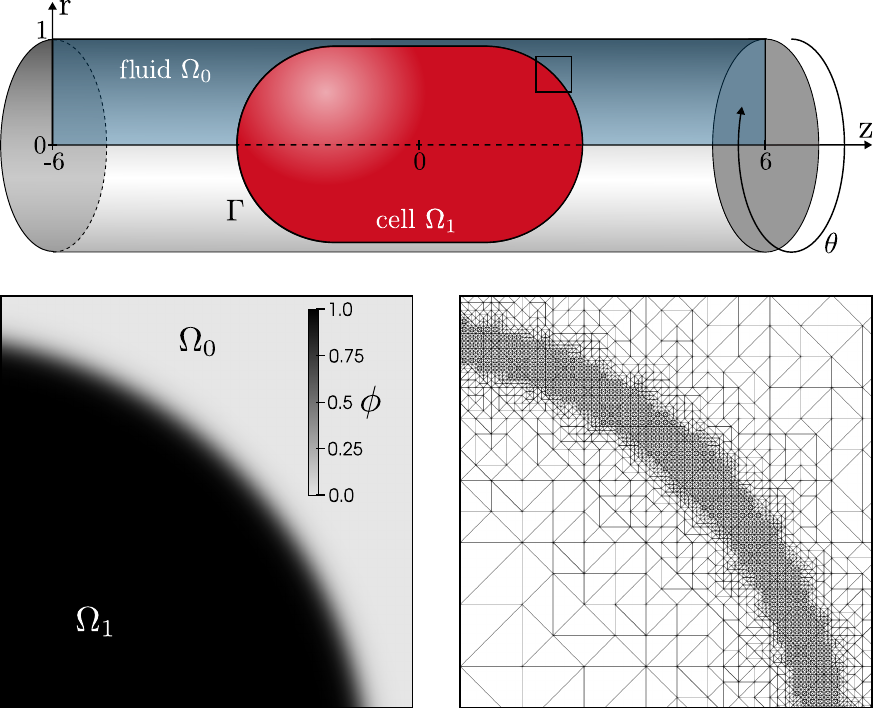}
    	\caption{
      {\bf-  Simulation Setup of a Cell in a Cylindrical Channel.
 Top}: Sharp-interface model with cell ($\Omega_{1}$) and fluid ($\Omega_{0}$) separated by interface $\Gamma$. Due to axisymmetry, only the blue rectangle is computed.
{\bf Bottom}: Diffuse-interface representation of the inset region by a phase field $\phi$ (left) and the corresponding adaptive mesh (right).} 	
	\label{fig:sketch}
\end{figure}

The spatial domain, denoted by \(\Omega\), is divided into the intracellular fluid domain \(\Omega_1\) and the exterior fluid medium \(\Omega_0\). Both domains are separated by the cell surface \(\Gamma\), representing the cell membrane and cortex, see Fig. \ref{fig:sketch} (top). 

We refer to the normal vector pointing from \(\Omega_0\) into \(\Omega_1\) as \(\mathbf{n}\), and use it to define the projection \(P_\Gamma\) onto \(\Gamma\) as $P_\Gamma \colon = I-\mathbf{n}\otimes\mathbf{n}$, where \(I\) denotes the identity matrix.
Differential operators on the surface \(\Gamma\) are defined as follows: 
\begin{align*}
	&\nabla_\Gamma \coloneqq P_\Gamma \nabla &&\text{surface gradient},\\
	&\nabla_\Gamma \cdot \coloneqq P_\Gamma : \nabla &&\text{surface divergence},\\
	&\Delta_\Gamma \coloneqq \nabla_\Gamma \cdot \nabla_\Gamma && \text{Laplace-Beltrami operator}.
\end{align*}
Note that applying these operators to a field variable requires the latter to be defined not only on $\Gamma$ but in a neighborhood of $\Gamma$. Throughout this article, all surface variables are assumed to be defined in this way.

We describe intracellular and extracellular flows in the whole domain $\Omega$ by the Stokes equations 
\begin{align*} 
	\nabla \cdot \left[\eta(\nabla \mathbf{u} + \nabla \mathbf{u}^T )\right]- \nabla p  &= \mathbf{f}_\sigma + \nabla_\Gamma \cdot (\delta_\Gamma S_{\text{visc}}) , \\
	\nabla\cdot\mathbf{u}&= 0, 
\end{align*}
where $\eta(  \textbf{x})$ denotes the fluid viscosity, both inside and outside the cell: $\eta(\textbf{x})= \eta_i$ for $\textbf{x}\in \Omega_i, i=0,1$. Moreover, $\mathbf{f}_\sigma$ and $S_{\text{visc}}$ denote the active surface tension and surface viscous stress, specified below, and  $\delta_\Gamma$ is a  
surface Dirac-delta distribution, which is defined as $\delta_\Gamma (x) = 0$ if $x \notin \Gamma$ and $\int_{\Omega} \delta_{\Gamma} \dx = |\Gamma|$, where $|\Gamma|$ is the area of any surface $\Gamma\subset \Omega$. 
We model the fluid dynamics using a single, continuous velocity field $\mathbf{u}$. The cell surface thus moves with this flow field. As a result, there is neither tangential slip nor fluid flow across the interface, leading to an intrinsically imposed no-slip condition on both sides of the cell surface.

Any flow and shape deformation along the cell surface is limited by the ability of the cortex to deform and remodel itself, such that accounting for the cortex mechanics is indispensable \cite{Kelkar.2020,FischerFriedrich.2016}. On the relevant time scales of cell division and cell migration, the rheology of the cortex is predominantly viscous  \cite{FischerFriedrich.2016}, which gives rise to a viscous surface stress $S_{\text{visc}}$, as introduced by Scriven \cite{Scriven.1960} as
\begin{align*}
	S_{\text{visc}}= (\eta_b-\eta_s)(\nabla_\Gamma \cdot \mathbf{u})P_\Gamma +\eta_s P_\Gamma(\nabla_\Gamma \mathbf{u} + \nabla_\Gamma \mathbf{u}^T)P_\Gamma,
\end{align*} 
where \(\eta_b\) and \(\eta_s\) are the bulk and shear viscosity of the surface.

The surface tension force $\mathbf{f}_{\sigma}$ is produced by force-generating molecules such as myosin motor proteins. As in previous literature  \cite{Mietke_PNAS_2019, Mietke_PRL_2019,AlandWittwer_2023}, we model the tension as a monotonically increasing Hill-function of the myosin concentration $c$ by \(\sigma(c) =\sigma_0\frac{2c^2}{c^2+c_0^2}\), where \(c_0\) is the  characteristic concentration and $\sigma_0>0$ a scaling factor. 
Consequently, the surface tension force can be formulated as 
\begin{align}
	{\bf f}_\sigma &= \nabla_\Gamma \cdot(\delta_\Gamma \sigma(c)P_\Gamma) = \delta_\Gamma H \sigma(c) {\bf n} + \delta_\Gamma \nabla_\Gamma \sigma(c),
\end{align}
where \(H= \nabla \cdot \mathbf{n}\) is the total curvature of the surface.
The first term on the right-hand side corresponds to the normal component of surface tension. Respectively, the second term is the tangential component, known as the Marangoni force, which arises from gradients of surface tension along the surface. It provides a force toward regions of high myosin concentrations, and is therefore responsible for the retrograde flows that drive amoeboid migration.

These equations are coupled to an advection-diffusion equation to describe the evolution of the myosin concentration \(c\), on \(\Gamma\):
\begin{align}\label{eq:c}
	\frac{{\rm d}c}{{\rm d}t} + c \nabla_\Gamma \cdot  \mathbf{u} - D \Delta_\Gamma c = 0,
\end{align}
where $ {{\rm d}}/{{\rm d}t}$ is the material derivative and $D>0$ the diffusion constant.
We omit more complex formulations of Eq.~\eqref{eq:c} that incorporate detailed binding/unbinding kinetics \cite{CallanJones.2013,AsanteAsamani.2024,Hawkins.2011}. 
We chose this simplification because these specific kinetics are not necessary to trigger the symmetry-breaking instability that triggers cell migration. 
Therefore, we use the simplest possible model, although these terms can be readily incorporated into the present framework.

\subsection{Non-Dimensionalization}
To reduce the number of model parameters, we non-dimensionalize the equations by rescaling length in units of the radius  \(R\) of the cylindrical channel, time in units of the diffusive timescale \(D/R^2\), and concentration in units of the characteristic concentration \({c_0}\) in the Hill function for the active stress.   
Then, dividing Eq.\eqref{eq:c}  by \({c_0R^2}/{D}\) yields the dimensionless concentration equation
\begin{align}
	\partial_t c + \nabla_\Gamma \cdot (c \mathbf{u})- \Delta_\Gamma c &= 0 &&\text{on }\Gamma. \label{eq:c nondim}
\end{align}
For the hydrodynamic equations, we base the scaling on the surface bulk viscosity $\eta_b$. Thus, we rescale pressure by $D\eta_b/R^3$ and introduce the surface viscosity ratio $\nu = \eta_s/{\eta_b}$ and the
{P\'eclet} number \(\Pe = \frac{\sigma_0 R^2}{D\eta_b}\) as the ratio between time scales of surface diffusion ($R^2/D$) and active flows
($\eta_b/\sigma_0$). We obtain 
\begin{align}
 &\hspace{-2em} -\nabla \cdot \left[ \frac{\eta R}{\eta_b} (\nabla \mathbf{u} + \nabla \mathbf{u}^T) \right] + \nabla p \nonumber \\
&\quad  = \Pe\, \delta_\Gamma \big( H \sigma(c) \mathbf{n} + \nabla_\Gamma \sigma(c) \big) \nonumber \\
&\qquad + \nabla_\Gamma \cdot \Big[ \delta_\Gamma (1 - \nu)\, P_\Gamma \nabla_\Gamma \cdot \mathbf{u} \nonumber \\
&\hspace{6.7em} + \nu\, \delta_\Gamma P_\Gamma (\nabla_\Gamma \mathbf{u} + \nabla_\Gamma \mathbf{u}^T) P_\Gamma \Big], 
\label{eq:NS 1 nondim} \\
&\qquad \nabla \cdot \mathbf{u} = 0,
\label{eq:NS 2 nondim}
\end{align}
with $\sigma(c) = \frac{2c^2}{c^2+1}$.
The dynamics of the system is thus governed by the three parameters $\Pe, \nu$ and $\eta R/\eta _b$. The values of the parameters used in our simulations are provided in Table \ref{tab:parameters} in the results section.

\subsection{Phase-field Ansatz}
To flexibly account for large deformations, topological transitions and wall contact, we introduce a phase-field (i.e. diffuse-interface) version of the above equations. In the phase-field model, an auxiliary field variable $\phi$ - the phase field - is introduced and used to indicate the bulk phases,
which can be arbitrary viscous fluids \cite{anderson1998diffuse}, as well as viscoelastic or elastic materials  \cite{mokbel2018_PF_FSI}. The phase-field function varies smoothly between the distinct values for the phases across the interface, resulting in a small but finite interface thickness $\epsilon$. 
Depending on the application of interest, phase-field methods may offer advantages over other interface-capturing methods. For example, they allow for unconditionally stable inclusion of surface tension  \cite{Aland_2014_time} and fully-discrete energy-stable schemes, see e.g.\cite{ChenAland,Gruen}. 

Here, we describe the geometry of the cell such that \(\phi(x)\approx 1\)  in $\Omega_1$ and \(\phi(x)\approx 0\) in $\Omega_0$  with a smooth transition between the two phases, see Fig. \ref{fig:sketch}(bottom). The phase field is initialized by $\phi=0.5  \big( 1+ \text{tanh}\left(   r/(\sqrt{2}\varepsilon) \right)\big)$, where $r$ is the signed distance to $\Gamma$, positive in $\Omega_1$. For complex interface geometries, $r$ can be computed numerically using redistancing algorithms  \cite{sussman1999efficient}. In our simulations, we employ a simple cigar shape (cylinder with two hemispherical caps, Fig.~\ref{fig:sketch}) which permits an explicit analytical expression for $r$.

After initialization, the phase field is advected with the fluid flow to capture changes in the surface geometry. This evolution is governed by the convective Cahn-Hilliard equation, which is solved over the entire computational domain \(\Omega\):
\begin{align}
	\partial_t \phi + \nabla \phi \cdot \mathbf{u} - \nabla \cdot (M \nabla \mu)  & =\frac{\beta}{\Delta t} \vert \nabla \phi \vert (V(0)-V(t)),  \label{CH-1}\\
	\mu + \varepsilon \Delta \phi - \varepsilon^{-1}W'(\phi) &=0 \label{CH-2},
\end{align}
where \(\mu\) denotes the chemical potential and \( W(\phi)=\phi^2(1-\phi)^2\) a double-well potential. The parameter \(\varepsilon>0\) describes the thickness of the interface region, and the mobility \(M>0\) governs the thermodynamically driven diffusion of the phase field.
In our approach, the diffuse interface merely approximates a sharp cell surface, and the phase‐field framework acts only to advect that interface consistently. Accordingly, interface motion is imposed by the velocity field rather than diffusion. We therefore choose $M\ll 1$, so that it primarily stabilizes and smooths the interface profile without shifting its position. 
The term scaled with constant $\beta>0$ serves as a volume-correction mechanism to relax the cell volume $V(t)=\int_\Omega \chi_{\phi>0.5} \dx$ back to its initial value $V(0)$. 
 In our simulations $\beta$ is chosen large enough such that volume loss is limited to roughly 1\%.
Note that the correction term compensates for the fact that the geometric volume enclosed by the  $\phi=0.5$ contour is not conserved under the Cahn-Hilliard dynamics, ensuring a more accurate control over the  domain occupied by the cell.

The phase-field representation can not only be used to describe the surface \( \Gamma = \{ {\bf x} | \phi({\bf x} )=0.5 \} \), but also to  {approximate} the Dirac-delta distribution \(\delta_\Gamma \approx \vert \nabla \phi \vert\).
Following the diffuse-interface approach  \cite{RaetzVoigt_2006}, the concentration equation \eqref{eq:c nondim} can be extended from the submanifold \(\Gamma\) onto \(\Omega\) as 
\begin{align}
	\partial_t (\vert \nabla \phi\vert c) + \nabla\cdot (\vert \nabla \phi\vert c \mathbf{u})- \nabla \cdot (\vert \nabla \phi\vert \nabla c) &=0 &\text{ in } \Omega. \label{eq:phase_field:conc}
\end{align}
See Appendix \ref{sec:diffuse interface derivation} for a derivation.
Even though the concentration equation suggests mass conservation, numerical discretization errors may lead to small deviations, which can accumulate to a significant mass loss over long simulation times. 
Therefore, we introduce the mass on the surface 
\[m(t) = \int_\Omega \vert \nabla\phi(t)\vert c(t) \dx \approx \int_{\Gamma(t)} c(t) \dx\] 
and add  a mass correction term
\[\frac{\alpha}{\Delta t} \vert \nabla \phi \vert (m(t)-m(0))\]
with some constant \(\alpha>0\) to the right-hand side of the concentration equation to ensure mass conservation.

Finally, we reformulate the momentum equation \eqref{eq:NS 1 nondim} in the phase-field formalism. The fluid viscosity is linearly interpolated between the distinct viscosities in the phases,
\begin{align}
    \eta(\phi)=(1-\phi)\eta_0 \frac{R}{\eta_b}+\phi\eta_1 \frac{R}{\eta_b}. \label{eq:viscosity}
\end{align}
As usual for diffuse-interface models of two-phase flow (e.g.  \cite{Aland_2014_time}), the constant surface tension term $\delta_\Gamma H\sigma {\bf n}$ can be reformulated to $3\sqrt{2}\sigma\mu\nabla\phi$, where the numerical
factor stems from the chosen double-well potential.
Moreover, we use the extended normal vector \( \mathbf{\tilde n} \coloneqq {\nabla \phi}/{\vert \nabla \phi \vert} \approx \mathbf{n}\) and the surface projection \( \tilde P_\Gamma \coloneqq I- \mathbf{\tilde n} \otimes \mathbf{\tilde n} \approx P_\Gamma \) to define diffuse-interface {approximations} to the surface differential operators, e.g. the surface gradient ${\tilde \nabla}_\Gamma \coloneqq \tilde P_\Gamma \nabla$ and surface divergence 
\({\tilde \nabla}_\Gamma \cdot \coloneqq \tilde P_\Gamma :\nabla \).
We obtain a diffuse-interface version of the surface stresses and forces
\begin{align*}
	\mathbf{f}_\sigma &= \Pe\left(3\sqrt{2}\sigma(c)\mu\nabla\phi + |\nabla\phi| \sigma'(c){\tilde \nabla}_\Gamma c\right), \\
	S_{\rm visc} &= (1-\nu){\tilde P}_\Gamma{\tilde \nabla}_\Gamma \cdot \mathbf{u} + \nu{\tilde P}_\Gamma({\nabla} \mathbf{u} + {\nabla} \mathbf{u}^T){\tilde P}_\Gamma 
\end{align*}
which enter the diffuse-interface hydrodynamics equation
\begin{align}
	-\nabla\cdot[\eta(\phi)(\nabla\mathbf{u}+\nabla\mathbf{u}^T)] + \nabla p &= \nabla\cdot(|\nabla\phi| S_{\rm visc}) +  \mathbf{f}_\sigma \label{eq:NS 1 phase-field} 
\end{align}
in $\Omega$.\\
Note that the first divergence operator on the right-hand side {of Eq.~\eqref{eq:NS 1 phase-field}} is not required to be a surface divergence, since it is applied to a tangential tensor. 

\subsection{Velocity Extension}
As we will see in the numerical tests, the active surface tension imposes strong tangential flows leading to regions of large tangential compression ($\nabla_\Gamma \cdot {\bf u}<0$) or stretching ($\nabla_\Gamma \cdot {\bf u}>0$). 
Due to incompressibility of the flow field, this goes along with the opposite deformation in the normal direction, i.e. tangentially stretched regions get compressed in the normal direction and vice versa. 
In numerical tests we find that such strong compressional or extensional flows in the normal direction tend to locally shrink or expand the thickness of the interface region, respectively, unless an extremely high, unphysical mobility is used.  
To eliminate this perturbation of the interface profile for reasonable mobilities, we instead advect the phase field with an auxiliary velocity field ${\bf v}$.
The idea is that ${\bf v}$ is constructed as an extension of the velocity ${\bf u}$ evaluated at the surface (i.e. at the 0.5-level set), which is constant in the direction ${\bf n}$ normal to the surface. 
Therefore, using ${\bf v}$ instead of ${\bf u}$ in the advective term, velocity differences across the diffuse interface are leveled out, which leads to a consistent advection of the complete diffuse interface region.
We propose to construct ${\bf v}$ by solving the additional equation for its components $(v_1,v_2,v_3)$,
\begin{align}
	\vert \nabla \phi \vert {v}_i -\vert \nabla \phi \vert {u}_i-\nabla \cdot [\vert \nabla \phi \vert \mathbf{\tilde n} ~\mathbf{\tilde n} \cdot \nabla {v}_i] &= {0}  \label{eq:phase_field:v_extended}
\end{align} 
on $\Omega$ and for $i=1,2,3$.\\
In the Appendix \ref{sec:asymptotic} we show by matched asymptotic expansion that this formulation converges to the following sharp-interface limit equations 
\begin{align}
	\mathbf{v} &= \mathbf{u}&\text{on }\Gamma,\label{eq:sharp_interface:u=v}\\
	\nabla \mathbf{v} \cdot \mathbf{n} &= 0 &\text{on }\Gamma.\label{eq:sharp_interphase:v'=0}
\end{align}
The obtained velocity field \(\mathbf{v}\) is used to replace the velocity \(\mathbf{u}\)  in the advection terms of both the Cahn-Hilliard equation and the concentration equation.

\subsection{Discretization}
To avoid numerical instabilities from capillary time step constraints, the six strongly coupled equations \eqref{eq:NS 2 nondim} - \eqref{eq:phase_field:conc} and \eqref{eq:NS 1 phase-field} - \eqref{eq:phase_field:v_extended}  are discretized by a monolithic linear semi-implicit time discretization  \cite{Aland_2014_time}. 
The proposed time discretization ensures mass conservation of regulating surface species on the discrete level as shown in Appendix \ref{sec:time}.

We solve the resulting system in each time step with a Finite Element method based on the Finite Element toolboxes DUNE  \cite{Sander_2020} and AMDiS  \cite{amdis2,VeyVoigt_2006,Witkowski_2015}.
An adaptive grid is employed to accurately resolve the phase field and surface forces with more details given in Appendix \ref{sec:space}. 
To make simulations more efficient, we assume that the cell shape and concentration field are axisymmetric. This assumption, which holds in particular for homogeneous cells in cylindrical channels, reduces computations effectively to a 2D domain from which the full 3D solution can be recovered, see Fig. \ref{fig:sketch}. Further details on the axisymmetric formulation can be found in Appendix \ref{app:axisymmetric}.

\section{Results}

To examine how cells determine their direction in confined geometries - such as microchannels or microcapillaries - we consider a cigar-shaped cell initially positioned inside a cylindrical channel (see Fig. \ref{fig:sketch} (top)). The cell is initialized with a length of 3.24 and a height of 1.84, placed within a channel of diameter 2. The channel length varies and is chosen sufficiently large to ensure that the cell is unaffected by the channel openings.
The interface thickness is selected to be sufficiently small such that the specific choice of $\varepsilon$ has little influence on the results; see Appendix \ref{sec:epsilon} for a more detailed discussion.
A nearly homogeneous cortical concentration field is prescribed with a base value of 1 and superimposed uniform noise centered on 0 and a width of 0.1. The channel walls are equipped with no-slip (${\bf v}=0$) and no-adhesion ($\phi=0, {\bf n}\cdot\nabla\phi=0$) boundary conditions. 
Unless otherwise specified the model and simulation parameters are as given in Tab.~\ref{tab:parameters}.

The chosen P\'eclet number corresponds to active cortical tension $\sigma_0$ on the order of $1$ mN/m as suggested from measurements in cells \cite{salbreux2012actin}. 
The critical {P\'eclet} number at which pattern formation occurs can be very roughly estimated from the linear stability theory in Ref.~\cite{Mietke_PRL_2019}, which however neglects the exterior fluid and channel walls. 
Based on our simulations, we identify a critical Péclet number of $\Pe \approx 20$ (see Appendix \ref{sec:peclet}).  
Starting simulations from random initial concentrations, we consistently find the emergence of dominant polar patterns of mode $1$ (i.e. high concentration at one cell end and low at the other).

The chosen ratio between cortical and fluid viscosity $\eta R/\eta_b = 1$ was selected to specifically study a balanced regime where the viscous contributions of the surrounding fluid and the cell cortex are comparable, clearly exposing the coupling mechanism necessary for cellular propulsion. 
While physically realistic values of cytoplasmic viscosity (see Tab.~\ref{tab:parameters}) suggest a much lower ratio, our results are expected to be robust as in both regimes single-spot myosin concentration patterns (1-modes) emerge \cite{Mietke_PRL_2019}.
To verify this, we performed additional representative simulations with a smaller fluid-to-cortical viscosity ratio ($\eta R/\eta_b = 0.25$). These confirmed that the cell consistently develops the same single-spot polarity, and that the overall influence of external cues on the symmetry-breaking mechanism remains robust.

\begin{table}[h]
\centering
\begin{ruledtabular}
\begin{tabular}{lll}
    time step &\(\Delta t\) & \(0.01\)\\    
    bulk grid size &\( h_{\text{bulk}}\) &  \(2.5\cdot 10^{-1}\)  \\ 
    interface grid size &\( h_{\text{int}}\) &\(5.52\cdot 10^{-3}\) \\  
    mobility &\(M\) &  \(10^{-3}\)\\
    interface thickness &\(\varepsilon\) & \(0.02\)\\
    volume correction &\(\beta\) & 0.01\\ 
    mass correction &\(\alpha\)& 0.25 \\
    surface viscosity ratio & $\nu = \eta_s/{\eta_b}$ & 1 \\
    fluid viscosity &  $\eta_0 R/\eta_b = \eta_1 R/\eta_b$ & $1$ \\
    {P\'eclet} number &$\Pe$ & 25 \\
    \hline\hline
    myosin diffusivity \cite{Maiuri.2015} & D & $100\,\mu$m$^2/$min\\
    viscosities \\
    - cortical \cite{FischerFriedrich.2016,Saha.2016} & $\eta_b, \eta_s$ & 1\,mN\,s/m \\
    - cytoplasmic \cite{Kalwarczyk.2011,Daniels.2006} & $\eta_1$ & $10^{-2}-10^1$ Pa\,s \\
    &$\eta_1 R/\eta_b$ & $10^{-4} - 10^{-1}$ \\    
    - exterior \cite{Forgacs.1998, Marmottant.2009, Guevorkian.2010, BlanchMercader.2017, PerezGonzalez.2019} & $\eta_0$ & $10^{-3}-10^{6}$ Pa\,s \\
    & $\eta_0 R/\eta_b$ & $10^{-5} - 10^{4}$ \\    
    cortical tension \cite{FischerFriedrich.2016,DizMunoz.2018} & $\sigma_0$ & 1\,mN/m\\
    cell/channel radius & $R$ & $10\,\mu$m \\
\end{tabular}
\end{ruledtabular}
\caption{Default model and simulation parameters (top) and experimental estimates of physical parameters (bottom)}.
\label{tab:parameters}
\end{table}

\begin{figure*}
    \centering
    \includegraphics[width=1\textwidth]{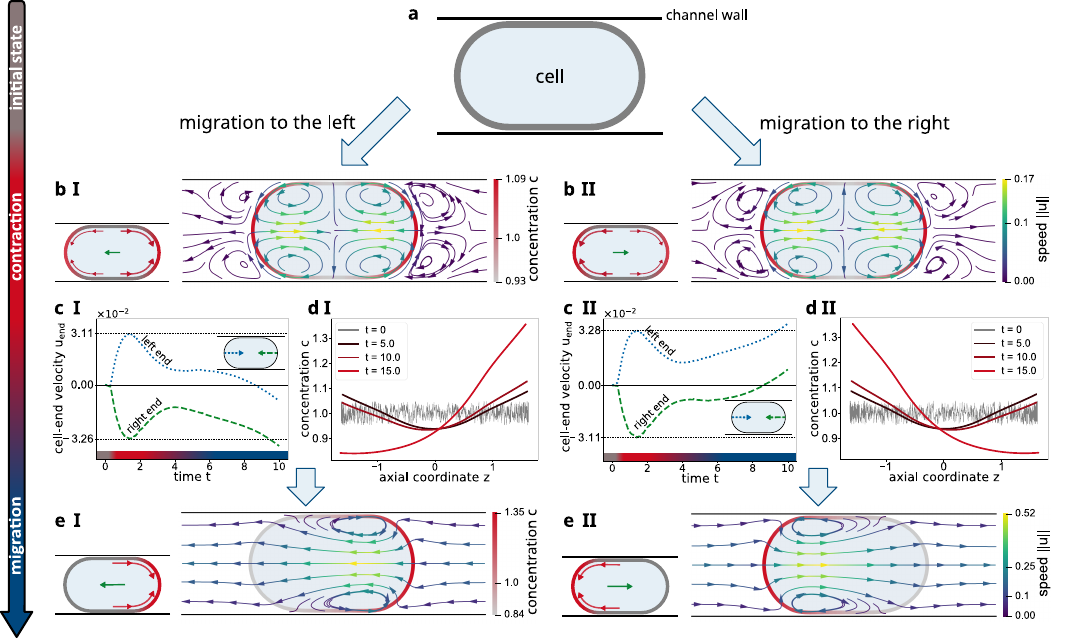}
    \caption{
\textbf{- Cell Migration via Spontaneous Symmetry Breaking in a Channel.}
\textbf{(a)} A cell is placed inside a channel and elongated relative to its relaxed shape. The surface concentration is initialized to be uniform with slight perturbations.
\textbf{(b)} Snapshot of the simulation at t = 1.36. The initial contraction, driven by surface tension, leads to the formation of two spots of high myosin concentration at the cell ends. Streamlines show the velocity field. In case I, the concentration and velocity are slightly higher on the right end; in case II, higher on the left end.
\textbf{(c)} Velocity at the cell ends over time. The blue dotted line corresponds to the left end and the green dashed line to the right end. Initially, the cell ends move towards each other as the cell contracts. However, the initial symmetry breaks spontaneously and the cell ends up migrating in one direction.
\textbf{(d)} Concentration profiles along the cell surface at different times. Early asymmetries in contraction lead to a buildup of concentration at one end, ultimately defining the cell rear and the direction of motion.
\textbf{(e)} Snapshot of the final state at t = 15. Over time, the asymmetries intensify, leading to full polarization and persistent migration to the left (case I) or right (case II).
Note that flow velocity arrows are displayed in the lab frame. Even though the vectors visually cross the cell surface, the no-penetration condition is fulfilled, as also the cell boundary moves at the local fluid velocity.}
    \label{fig:Spontaneous}
\end{figure*}

\subsection{Spontaneous Symmetry Breaking}
Studying the polarization process experimentally is challenging due to the small spatial scales and complex interplay of physical forces involved. Our modeling approach allows us to overcome these limitations and gain mechanistic insight into how polarization and migration emerge in non-adherent cells. To begin, we consider a single cell confined within a channel, without any external cues, in order to isolate the intrinsic mechanisms driving polarization and movement.

As shown in Figure \ref{fig:Spontaneous}, the cell initially rounds up (i.e. reduction in length and increase in width) and adapts to the channel. 
Due to the properties of the phase-field model -- specifically, the boundary condition $\phi = 0$ at the channel walls and the resulting phase-field profile -- a thin liquid film remains between the cell and the wall.
The thickness of this film scales with the interface thickness $\varepsilon$. A systematic validation confirms that, for sufficiently small $\varepsilon$, the results are insensitive to its precise value (see Appendix \ref{sec:epsilon} for details).

\begin{figure*}
    \centering
    \includegraphics[width=0.935\textwidth]{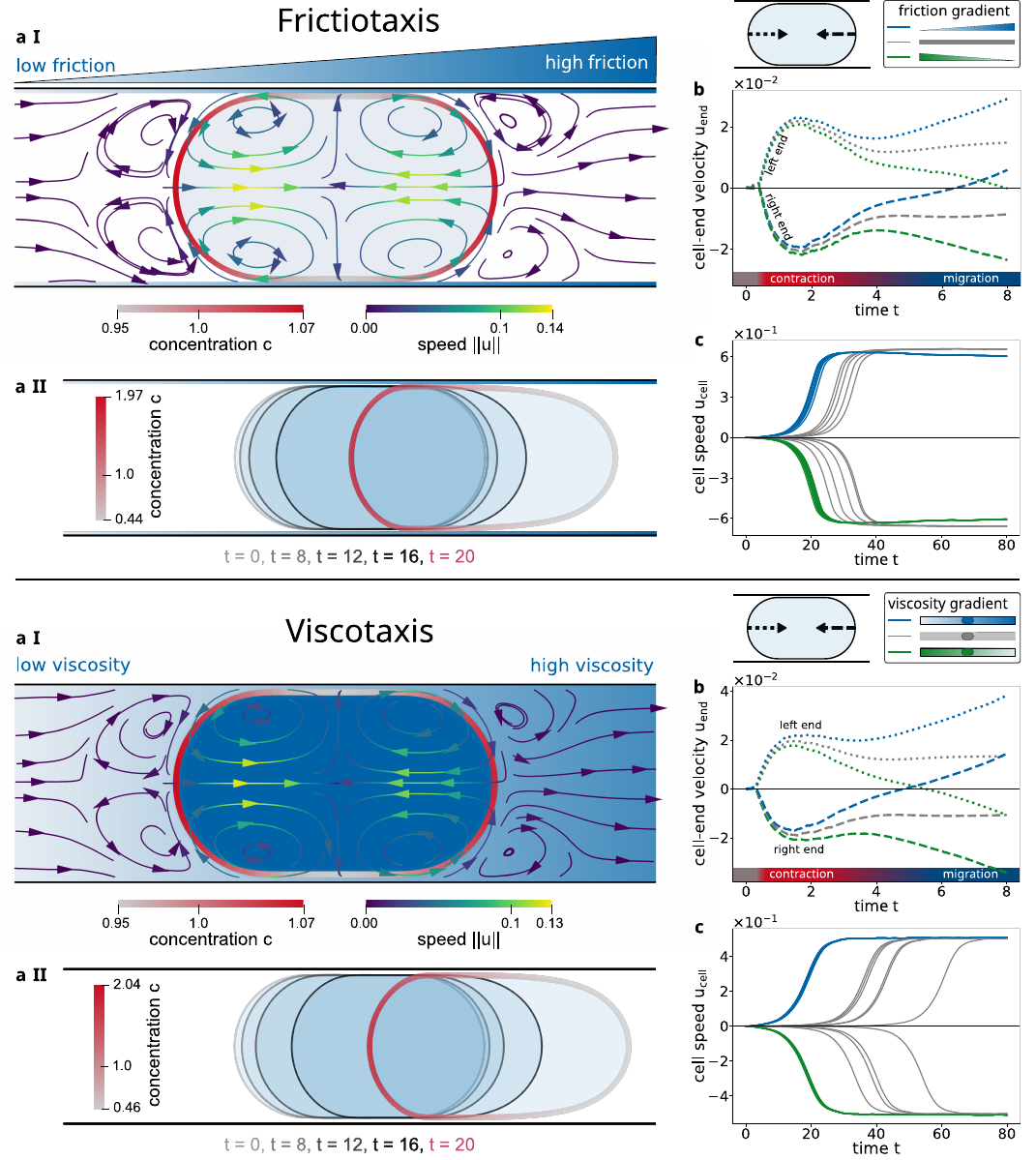}
    \caption{\textbf{- Guidance of Cell Migration by Friction and Viscosity Gradients.}
A cell is initialized in a channel with either a friction gradient on the channel wall (top panel) or a viscosity gradient in the surrounding fluid (bottom panel). The internal viscosity of the cell remains uniform throughout. 
\textbf{(a I)} 
A snapshot of the cell at t = 1.65 shows an asymmetric formation of high-concentration spots at the cell ends. Streamlines indicate asymmetric velocity fields, with higher velocity and concentration observed on the low-friction or low-viscosity side, respectively. 
\textbf{(a II)} Outlines of the cell at t = 0, 8, 12, 16, and 20. The cell undergoes asymmetric contraction and progressively migrates towards higher friction or viscosity. At t = 20 the cell is fully polarized with a single spot of high concentration at the rear of the cell.
\textbf{(b)} Velocity at the cell ends over time for varying friction and viscosity gradients. In both conditions, contraction and velocity are greater on the low-friction or low-viscosity side. For uniform environments (gray curves), cell end velocities remain nearly symmetric.
\textbf{(c)} Cell velocity over time for ten different initial conditions each for three cases: uniform friction/viscosity (gray) and a gradient to the left (green) or right (blue) of the channel. In uniform environments, symmetry is broken spontaneously, and the cell moves either left or right with equal probability. Under gradient conditions, cells consistently migrate toward regions of higher friction or higher viscosity.
Parameters: $k_{\text{right}}= 1000, k_{\text{left}}= 0, \Pe = 22, \text{ and } \eta_{\text{right}} = 1, \eta_{\text{left}} = 0.25, \Pe = 15$
}
    \label{fig:frictio_visco}
\end{figure*}

As the cell contracts, two regions of high myosin concentration emerge at the poles.
Although this state is nearly symmetric, small asymmetries - stemming from noise in the initial myosin concentration - gradually amplify over time through a self-reinforcing process driven by the Marangoni force. Because the myosin concentration at one end is slightly higher, there is a net flow of myosin towards that region. This net flow eventually causes the peak at one side to outcompete the other, pulling in all the available myosin and breaking the initial symmetry. This process of spontaneous symmetry breaking forms a single spot of high myosin concentration with equal probability at either end of the cell.

Eventually the cell fully polarizes, Fig. \ref{fig:Spontaneous}(d,e). The asymmetric distribution of myosin generates a constant force toward the high-concentration region, inducing a retrograde flow (in the cell’s frame of reference) of the cell cortex. 
This flow interacts frictionally with the confining walls, through shear stresses transmitted by the thin liquid layer between the cell surface and the channel wall,
allowing the cell to transmit tangential forces to its surrounding. As the rear contracts and the cortex flows backward, its friction against the channel walls propels the cell body forward. At the same time, the cytoplasm's incompressibility leads to a recirculating flow pattern within the cell, Fig.~\ref{fig:Spontaneous}(e).

Note that the frictional interactions with the channel walls can arise through two distinct mechanisms. First, they may result from the close proximity with a no-slip wall, which inhibits tangential motion of the cell surface, either by direct contact or by strong viscous shear forces that penalize velocity differences between the cell and the near boundary.
In contrast to these hydrodynamic effects, friction can also be modeled as an explicit force opposing fluid motion. This latter approach is employed to investigate the recently observed phenomenon of friction-guided cell migration in the next section.

\subsection{Frictio- and Viscotaxis: Cells Migrate Up Friction and Viscosity Gradients}
Cells rarely migrate in a purely random manner; rather, their movement is often directed by external clues - for example, in response to nutrient sources or during the wound healing process. While migration guided by chemical gradients - known as chemotaxis - is well studied  \cite{Paluch.2016,Shellard.2020,Condeelis.1990,Shi.2013,SenGupta.2021}, cell migration guided by mechanical cues remains relatively less explored. In particular for adhesion-independent migration in channels, recent work found that cells can follow gradients of friction - a phenomenon that was called frictiotaxis  \cite{Frictiotaxis.2025}. Here, we explore this new mode of directed migration with our active surface model. Moreover, we also consider guidance by gradients of viscosity of the surrounding medium.

Using in vitro microchannels, Ref.\cite{Frictiotaxis.2025} showed that cells tend to migrate persistently up a friction gradient - that is, toward regions of higher wall friction. The authors also proposed a simple one-dimensional model to explain how cells polarize towards higher frictions.
Here we extend this analysis by investigating frictiotaxis through our 3D model, which accounts for changes in cell shape and captures the fluid flows not just in the cortex but also in the cytoplasm and the extracellular medium.

In our axisymmetric 3D setup, the friction gradient is incorporated by introducing an additional friction force $\mathbf{f}_{\rm fric} = -\chi_{\rm wall} k(\rm{z})~ \mathbf{u}\cdot \hat{\mathbf{z}}~\hat{\mathbf{z}} $ 
to the right-hand side of the force-balance  equation \eqref{eq:NS 1 nondim}. 
This force is oriented in the axial direction $\hat{\mathbf{z}}$ opposing the axial velocity $\mathbf{u}\cdot \hat{\mathbf{z}}$, scaled with a cell-wall friction coefficient $k(\rm{z})$ 
which varies linearly across the channel ranging from $k_{\text{left}}$ on its left to $k_{\text{right}}$ on its right end. 
The force is localized by the characteristic function $\chi_{\rm wall}$, which equals 1 only within a thin friction layer of thickness 0.05 and 0 elsewhere. This ensures that the friction force is integrated solely over a narrow region adjacent to the channel wall. 

As shown in Fig.~\ref{fig:frictio_visco}(top), the cell develops regions of high myosin concentration at both ends at early times. However, the symmetry in the flow field is broken by the presence of the friction gradient, see Fig.~\ref{fig:frictio_visco}(a I). 
The lower friction on one side drives two key processes. First, it promotes faster contraction. Second, it accelerates the surface transport of myosin toward that same side. Both of these effects lead to a faster accumulation of myosin on the low-friction side.
This process is self-reinforcing due to the Marangoni force, leading to the cell polarization. As a result, the cell migrates up the friction gradient driven by the resulting retrograde flows, Fig.~\ref{fig:frictio_visco}(a II). 
A closer examination of the velocity at both cell ends reveals that this symmetry breaking occurs very early in the contraction phase  (Fig.~\ref{fig:frictio_visco}(b)). The cell speed \( u_{cell} = \int_\Omega \phi u_z \dx / V(0) \) continuously increases on longer time scales until a maximum velocity is reached, which then slowly decays due to the constantly increasing friction, Fig.~\ref{fig:frictio_visco}(c). Note that this decay does not occur when friction is uniform (Fig.~\ref{fig:frictio_visco}(c, gray)).

Migration up the friction gradient is consistently observed for many different initializations of the initial conditions, Fig.~\ref{fig:frictio_visco}(c), illustrating the robustness of this mechanism.
Hence, our simulations confirm the previous 1D model results \cite{Frictiotaxis.2025}, highlighting that the frictiotaxis mechanism remains robust when including the effects of cell-shape changes and 3D hydrodynamics.

Depending on the wall coating, previous works reported values of the cell-wall friction coefficient in the range $10^3 - 10^7$\,Pa\,s/m \cite{Frictiotaxis.2025,Bergert.2015}. Moreover, Ref.\cite{Frictiotaxis.2025} experimentally demonstrated frictiotaxis in channels with a two-fold increase in friction along a channel of $\sim 100\,\mu$m. 
Translating our simulation units into physical units using the parameter estimates in Tab.~\ref{tab:parameters}, we impose a strong friction gradient of $5\cdot 10^8$\,Pa\,s/m along a chanel of $120$ $\mu$m.

Another emerging area of research is viscotaxis - the directed migration of cells in response to spatial variations in environmental viscosity  \cite{Shellard.2020,Liebchen.2018}. In this study, we investigate how a viscosity gradient in the surrounding medium influences cell behavior. To incorporate this effect, we modify the previously uniform viscosity, described in Eq.~\eqref{eq:viscosity}, by introducing a linear gradient in extracellular viscosity that transitions from $\eta_{\text{left}}$ at the left end to $\eta_{\text{right}}$ at the right end of the channel. The intracellular viscosity remains constant throughout and we don't include the additional friction term.
Note that the left/right gradient in extracellular viscosity naturally extends to the thin liquid film between cell boundary and channel wall.

Simulation results (Fig.~\ref{fig:frictio_visco}(bottom)) show that cells consistently migrate toward the region of higher viscosity. The underlying mechanism closely resembles that of frictiotaxis: lower extracellular viscosities enable faster cortical flows, leading to faster accumulation of myosin. 
This initial asymmetry strengthens over time, resulting in cell polarization and directed migration up the viscosity gradient as observed consistently across a range of simulations with different initializations of the initial conditions, Fig.~\ref{fig:frictio_visco}(c).

Translating our simulation units to physical units based on the parameter estimates in Tab.~\ref{tab:parameters}, the viscosity gradient imposed in our simulations is of the order of $1$\,Pa\,s/$\mu$m.
As previously discussed, these viscosity values are considerably higher than those found in typical physiological conditions. 
However, such high values could be relevant for cell migration within viscous tissues. 
Tissue viscosities are far greater than those of liquid biological media; they are often in the range $10^3 - 10^6$ Pa s/$\mu$m \cite{Forgacs.1998, Marmottant.2009, Guevorkian.2010, BlanchMercader.2017, PerezGonzalez.2019}, which can give rise to viscosity gradients potentially much larger than the one we utilized.

\subsection{Rheo- and Barotaxis: Cells go with the flow}

\begin{figure*}
    \centering
    \includegraphics[width=1\textwidth]{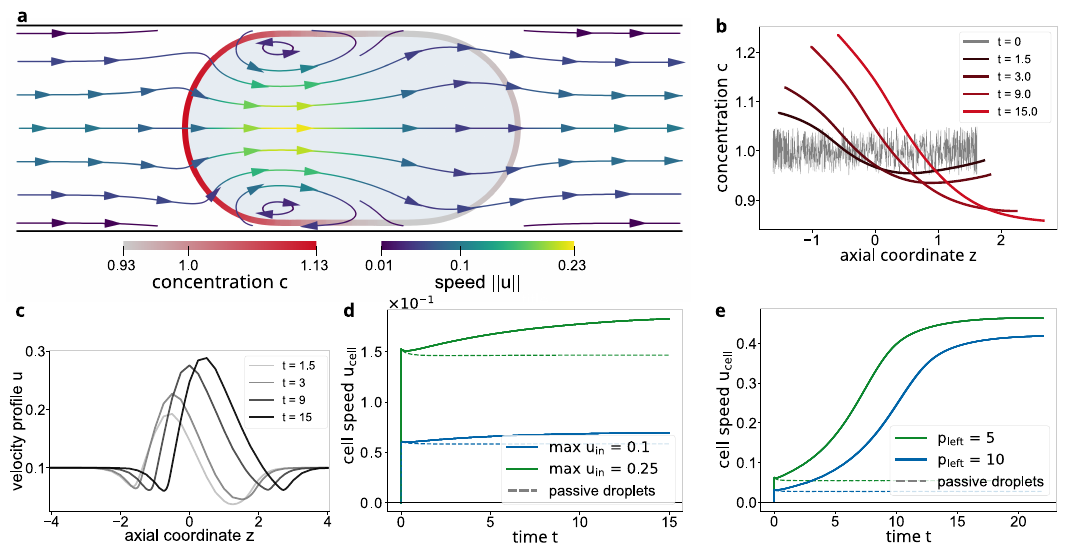}
    \caption{\textbf{- Guidance of Cell Migration by Inflow and Pressure Gradients.} (a-d) Rheotaxis: A cell in a channel with no-slip walls is exposed to an imposed inflow from left to right. 
    \textbf{(a)} A snapshot of the cell and surrounding flow field at t = 3.0 shows the development of a high-concentration spot at the rear, accompanied by retrograde cortical flows. Streamlines indicate flow patterns inside and around the cell. 
    \textbf{(b)} Concentration along the cell surface at different times, showing the emergence and strengthening of polarity. 
    \textbf{(c)} Velocity profiles along the channel centerline at multiple times, revealing asymmetric propulsion in the direction of the imposed flow. 
    \textbf{(d)} Cell velocity over time for two inflow magnitudes ($\rm {u}_{\rm in}$ = 0.1 and 0.25), comparing active cells (solid) and passive droplets (dashed); active cells consistently move faster.
    \textbf{(e)} Barotaxis: Cell velocity over time for two pressure drops ($\rm{p}_{\rm left}$ = 5 and 10; $\rm{p}_{\rm right}$ = 0), comparing active cells (solid) and passive droplets (dashed). Active cells migrate toward the low-pressure end and reach higher speeds than passive droplets.
    }
    \label{fig:rheo_baro}
\end{figure*}

In many physiological and experimental settings, cells experience fluid flows that influence their polarization and migration. Flow-induced navigation, or rheotaxis, occurs when an externally imposed flow field interacts with intrinsic polarization mechanisms. 
To simulate this, we impose an inflow velocity $\rm  u_{\rm in}$ at the open channel boundaries. Under no-slip wall conditions - mimicking typical microfluidic experiments - cells consistently migrate in the direction of the flow. 
This behavior closely resembles that of passive droplets, which we model here by setting a uniform surface tension ($\sigma(c) \equiv 1$) and thereby decoupling the surface tension from the myosin concentration. 

The imposed flow pushes the cell forward while its surface remains stuck to the channel walls, thus generating a retrograde cortical flow in the cell’s frame of reference. This retrograde flow, together with enhanced contraction on the upstream side and reduced contraction downstream, drives myosin accumulation at the rear, reinforcing polarization aligned with the flow (Fig.~\ref{fig:rheo_baro}(a)). Active migration thus adds to the passive advection, enabling the cell to move faster than a passive droplet (Fig.~\ref{fig:rheo_baro}(d)).

Using the parameter values in Tab.~\ref{tab:parameters}, the used inflow parameters correspond to velocities on the order of 1\,$\mu$m/min. The resulting cell speeds are of the same order, consistent with experimental observations. 

In confined tissues and microfluidic setups, pressure gradients are another common physical cue that can bias cell motion - a phenomenon known as barotaxis. Unlike rheotaxis, where velocity is imposed directly, here the driving force is a difference in hydraulic pressure across the channel. In vitro experiments have shown that many cell types preferentially move toward the low-pressure branch in bifurcated channels, with cell polarity strongly correlated to directional bias \cite{LennonDumenil.2021,Moreau.2019}. 

We simulate barotaxis by imposing a pressure gradient: a positive pressure at the left boundary and zero at the right, thereby inducing flow toward the low-pressure end - analogous to rheotaxis. Again, retrograde flow and rearward contraction both promote myosin accumulation at the rear, producing sustained forward propulsion, closely mirroring the results of rheotaxis. This behavior is robust across different initializations of the initial conditions Fig.~\ref{fig:rheo_baro}(e)). 
Using the parameter values in Tab.~\ref{tab:parameters}, the imposed pressure gradient corresponds to approximately 0.07\,Pa/$\mu$m. 
This gradient would result in passive flow speeds below $1\,\mathbf{\mu\text{m}/\text{s}}$, which seem representative of the slow flows observed in confined biological environments.

In rheotaxis, the fluid speed is constrained by the prescribed velocity at the channel openings, which prevents migrating cells from propelling the surrounding incompressible fluid.
In contrast, during barotaxis, the fluid velocity at the channel openings is not restricted. This allows migrating cells to accelerate the entire fluid in the channel, resulting in a substantially amplified speed difference between active cells and passive droplets (Fig.~\ref{fig:rheo_baro}(e)). 

\subsection{Topotaxis: Cells Migrate Away from Confinement}

\begin{figure*}
    \centering
    \includegraphics[width=1\textwidth]{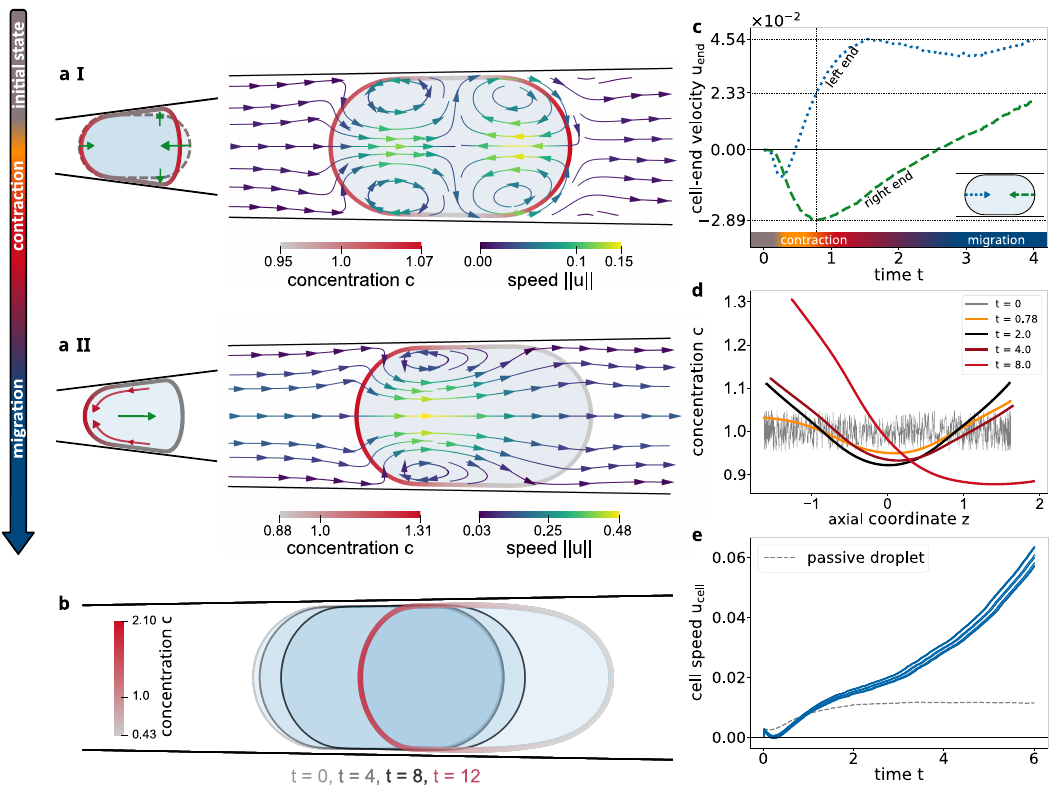}
    \caption{\textbf{- Guidance of Cell Migration by Environmental Topography (Topotaxis).}
A cell is initialized inside a slanted channel with a wider opening on the right.
\textbf{(a I)} Snapshots of the cell at t = 0.75. Initially, surface tension causes the cell to conform to the geometry of the slanted channel (see sketch), resulting in increased concentration and velocity on the right side of the cell. 
\textbf{(a II)} At later time (t = 8), the cell polarization has reversed and  persistent contraction on the left drives net migration to the right. Retrograde flow near the channel wall transports myosin toward the rear, reinforcing directional polarization. 
\textbf{(b)} Cell conformation shows increasing asymmetry over time  (t = 0, 4, 8 and 12), indicating progressive polarization and migration.
\textbf{(c)} Velocity at the cell ends over time. After an initial conforming-to-the-channel phase (orange) and contraction phase (red), the cell migrates consistently to the right (blue).
\textbf{(d)} Concentration profiles along the cell surface at different times. Initially, two peaks emerge, with the right side being more pronounced (t = 0.78, orange). After t = 2.0 the pattern shifts and polarizes with higher concentration on the left side.
\textbf{(e)} Cell velocity over time for ten different initializations of the initial conditions shows that active cells consistently move away from confinement, with a velocity much faster than a passive droplet with constant surface tension (dashed line).}
    \label{fig:Topotaxis}
\end{figure*}

In addition to sensing biochemical signals, cells are highly responsive to mechanical constraints and spatial confinement. Changes in geometry of the extracellular medium - whether due to tissue structure, extracellular matrix composition, or physical barriers - can significantly influence how cells move  \cite{Reversat.2020,Shellard.2020,SenGupta.2021}. Given the ubiquity of such conditions in vivo, it is important to examine how varying degrees of confinement shape migratory responses.

Rather than considering complex three-dimensional geometries, we focus here on a simple slanted (conical) channel with an opening angle of approximately 1\textdegree to demonstrate that even subtle geometric changes can strongly influence cell behavior. 

Initially, a cigar-shaped cell is placed in a slightly slanted channel. Symmetry breaking occurs in two stages: In the first stage, the cell adapts its shape to the channel's geometry and contracts more on the side with greater available space, as it can extend further in the perpendicular direction there. This expansion leads to a localized increase in molecular concentration and flow velocity on the less-constricted side (Fig.~\ref{fig:Topotaxis}(a I)), until the cell has adapted to the channel geometry at $t \approx 0.78$. This dynamics is also observed in simulations with passive droplets ($\sigma=1$, results not shown). 

Despite the slight asymmetry in concentration resulting from the cell’s initial adaption to the channel geometry, surface tension remains largely homogeneous in active cells at this early stage. Consequently, the cell, much like a passive droplet, seeks to minimize its surface area and moves towards the wider region of the channel. During this movement towards the channel opening, the cell cortex, experiences frictional drag from the nearby channel walls, causing it to move slower than the cell body. The difference in velocity constitutes a retrograde flow in the cell’s frame of reference transporting myosin to the more-constricted end. At $t \approx 2$, the increasing myosin concentration on the constricted end surpasses the concentration on the less-constricted side (Fig.~\ref{fig:Topotaxis}(d)). 

This behavior is consistent with simulations of both active cells and passive droplets. However, while passive droplets are driven solely by homogeneous surface tension, active cells migrate significantly faster due to the Marangoni forces created by the myosin concentration gradient. This reinforces the asymmetry and accelerates the migration process to significantly larger speeds (Fig. \ref{fig:Topotaxis}(e)).

These findings suggest that directed migration into constricted environments, as reported in some studies \cite{Reversat.2020,Park.2018}, requires additional mechanisms beyond those modeled here.

\begin{figure*}
    \centering
    \includegraphics[width=1\textwidth]{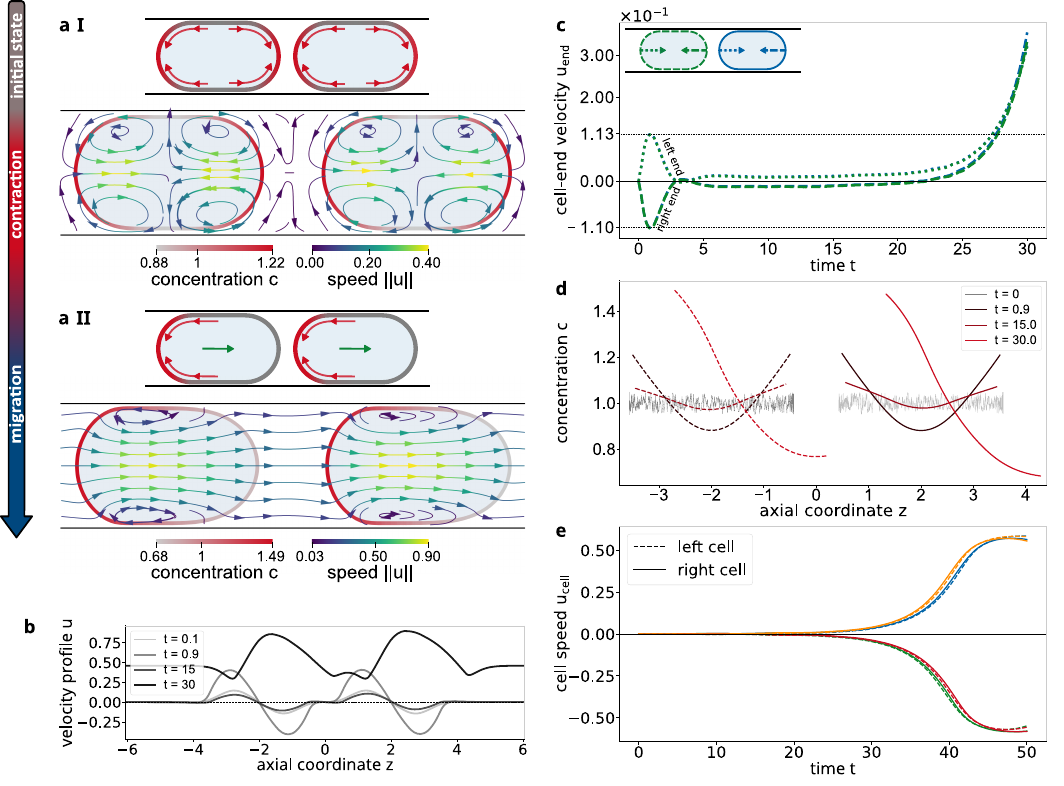}
    \caption{\textbf{- Coordination of Cell Migration via Hydrodynamic Interaction.}
Two cells are initialized side by side within a channel, each with a nearly uniform surface concentration and elongated shape. There is no direct physical contact between the cells - interaction occurs purely through the surrounding flow field.
\textbf{(a I)} At an early time (t = 0.9), both cells contract due to surface tension and behave nearly independently. Each cell induces its own flow, which partially cancel each other out between the cells and reduces velocities on the inner sides.
\textbf{(a II)} At t = 30, one cell is slightly more polarized and drives the neighboring cell forward. Both cells exhibit rearward localization of concentration and migrate in the same direction.
\textbf{(b)} Velocity profiles along the channel centerline over time. Early profiles show near-zero velocity between the cells (at z = 0) due to mutual cancellation of the opposing flows. 
\textbf{(c)} Velocity at the cell ends over time. Initially, both cells (blue and green) contract inward asymmetrically. The differences between the cell ends (dotted left, dashed right) of each cell are larger than differences between the two cells. Around t = 20, the cells begin to mutually migrate in one direction.
\textbf{(d)} Concentration profiles along the surface of both cells (dotted left, solid right). Early contraction causes accumulation at both cell ends. Over time, the initial concentration differences in each cell decrease, and the profile transitions into a single peak at the rear, indicating polarization.
\textbf{(e)} Cell velocity over time for four different initial conditions. In all cases, the two cells ultimately migrate in the same direction, illustrating the robustness of hydrodynamically mediated coordination.
Parameters:  $\Pe = 20$ }
    \label{fig:2cells}
\end{figure*}

\subsection{Hydrodynamic Interactions Coordinate Motion of Nearby Cells}

Cells are typically not isolated in biological organisms but reside in close proximity to neighboring cells, often forming cooperative groups such as cell trains \cite{Wortel.2024}. It is therefore of interest to investigate how polarization and migration are influenced by the presence of a neighboring cell. To this end, we consider a minimal system consisting of two identical cells placed side by side within a homogeneous narrow channel.
The two cells are not connected by adhesive interactions or signaling pathways but are coupled solely through the shared velocity field in the surrounding medium, Fig.~\ref{fig:2cells}(a). 

During the initial contraction phase, the cortical flows generated by each cell interfere with one another, leading to a partial cancellation of the flow field in the region between them, Fig.~\ref{fig:2cells}(b). Despite this, the cells undergo symmetry breaking, with one of the two cells breaking symmetry earlier than the other. As a result, the cortical flows are slightly stronger in the cell where symmetry broke earlier, which hence achieves a stronger polarity earlier, Fig.~\ref{fig:2cells}(d). 
As this small difference amplifies over time, the more strongly polarized cell begins to move, and - through their coupling via the incompressible fluid between them - it drives the other cell to move in the same direction.
The resulting pair migration proceeds at a speed comparable to that of a single cell, suggesting that mutual hydrodynamic interactions facilitate directional coherence, Fig.~\ref{fig:2cells}(e).

This minimal two-cell setup provides a foundation for exploring more complex collective behaviors. In particular, it raises intriguing questions about the outcome of interactions between already polarized cells approaching one another from opposite directions - an avenue we propose for future study.

\section{Conclusion}

\begin{table*}
    \begin{ruledtabular}
        \begin{tabular}{ >{\centering\arraybackslash}m{2.25cm} 
                         >{\centering\arraybackslash}m{3.1cm} 
                          >{\centering\arraybackslash}m{3.5cm} 
                          >{\centering\arraybackslash}m{3cm} 
                          >{\centering\arraybackslash}m{1.5cm} 
                          >{\centering\arraybackslash}m{2cm} }
\textbf{Migration mode}  & \textbf{Schematic Representation} & \textbf{Model Modifications} & \textbf{Migration Direction}  & 
\textbf{Passive Droplets} & \textbf{References}\\
\hline
\textbf{Frictiotaxis} &  \begin{minipage}{\linewidth}
      \includegraphics[width=3cm]{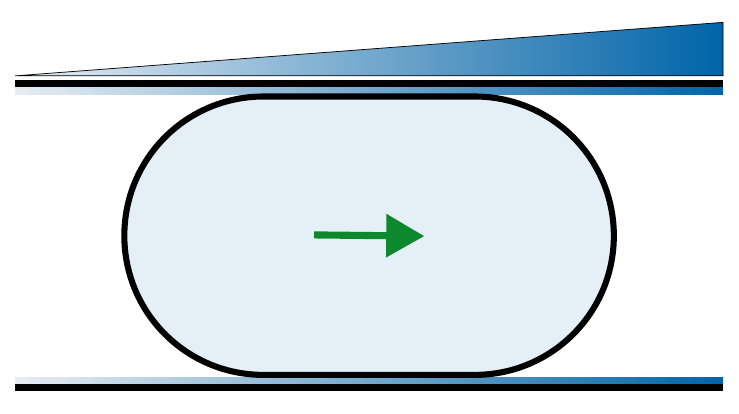}
    \end{minipage}  
    &  Additional friction force term in the momentum  equation \eqref{eq:NS 1 nondim} & Towards high friction & $\times$ & \cite{Frictiotaxis.2025} \\ 
\textbf{Viscotaxis} & \begin{minipage}{\linewidth}
      \includegraphics[width=3cm]{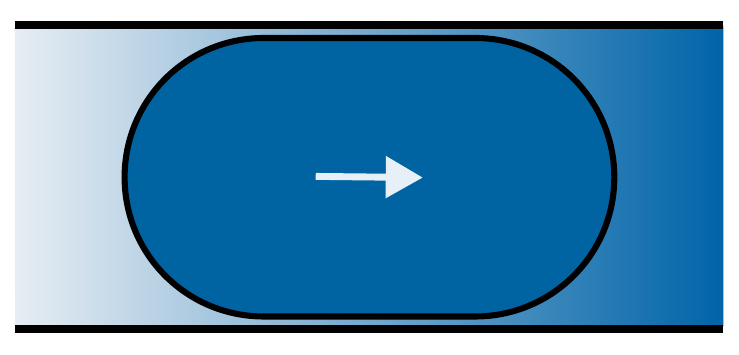}
    \end{minipage}  
& Introduction of a gradient in extracellular viscosity & Towards high viscosity & $\times$ & \cite{Liebchen.2018} \\
\textbf{Rheotaxis} & \begin{minipage}{\linewidth}
      \includegraphics[width=3cm]{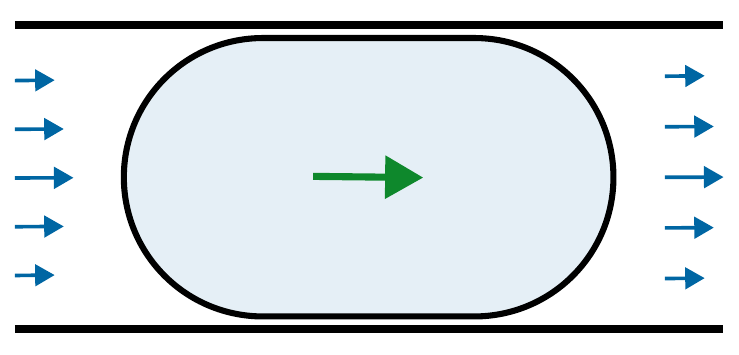}
    \end{minipage}   & Inflow at the channel openings & With the flow & $\checkmark$ & \cite{Bergert.2015,Otto.2015}\\ 
\textbf{Barotaxis} & \begin{minipage}{\linewidth}
      \includegraphics[width=3cm]{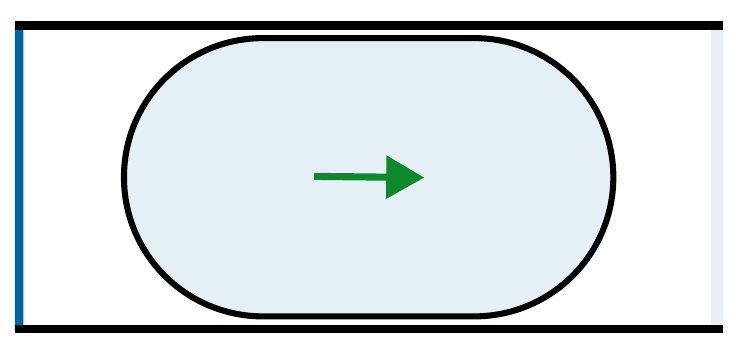}
    \end{minipage}   & Additional pressure boundary conditions & Towards low pressure & $\checkmark$ & \cite{LennonDumenil.2021,Moreau.2019} \\ 
\textbf{Topotaxis} & \begin{minipage}{\linewidth}
      \includegraphics[width=3cm]{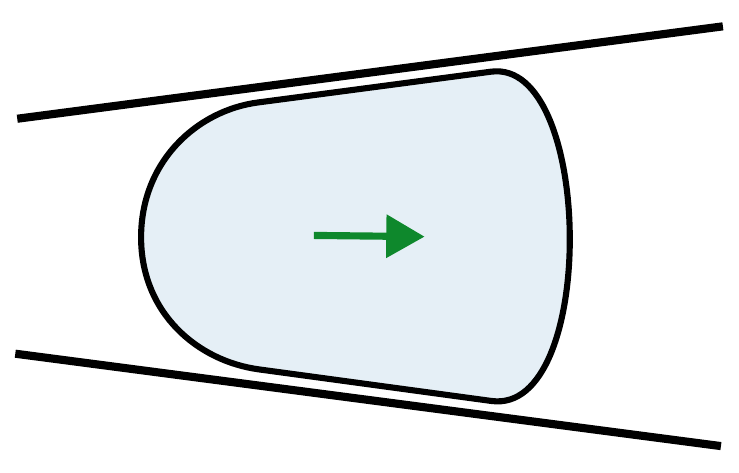}
    \end{minipage}   & Alteration of the channel geometry & Away from confinement & $\checkmark$ & \cite{Reversat.2020,Park.2018} \\
        \end{tabular}
    \end{ruledtabular}
    \caption{Modeling Modifications and Behaviors in Different Modes of Directed Cell Migration}
    \label{tab:conclusion_table}
\end{table*}

We have introduced a phase-field model to study the motility of non-adherent, deformable cells guided by a contractile instability of the cell cortex. The model couples surface and bulk hydrodynamics to surface flow of a diffusible species (myosin), which generates an active contractile force. It accounts for surface viscosity and includes a stabilizing auxiliary velocity field, derived through asymptotic analysis, to maintain robust interface dynamics under strong cortical deformations.

Through simulations, we systematically explored mechanisms guiding non-adherent cell migration. In the absence of external cues, confined cells break symmetry spontaneously via contractile activity, forming a single myosin-rich spot and migrating persistently through retrograde flow. Introducing gradients in the environmental properties revealed distinct taxis behaviors (see Table \ref{tab:conclusion_table}): frictiotaxis and viscotaxis arise from gradients in wall friction and extracellular viscosity, respectively, both directing migration toward higher-resistance areas. Rheotaxis, induced by external flow, aligns polarization with the flow direction, causing active cells to move faster than passive droplets. Barotaxis, driven by pressure differences, directs movement toward low-pressure zones. Topotaxis, triggered by gradients in environmental geometry, leads to migration away from constriction, resembling passive droplet behavior. Finally, we examined cell pairs, showing that hydrodynamic coupling induces collective migration of both cells in the same direction. Because of their lack of focal adhesion, both with the environment and with other cells, non-adherent cells are thought to migrate mostly as single cells. Our findings reveal that hydrodynamic interactions could lead to collective amoeboid migration, which was observed in recent experiments  \cite{Wortel.2024,Pages.2022}.

We show that mechanical cues, such as friction, viscosity, pressure, flow, and confinement, can each independently guide non-adherent cell migration. Moreover, even weak hydrodynamic interactions enable coordinated motion in multi-cell systems. Notably, passive and active responses differ under specific conditions, emphasizing the role of the contractile instability of the cell cortex in shaping stimulus-specific behaviors.

Translating our simulation units into physical units using the parameter estimates in Tab.~\ref{tab:parameters}, the characteristic time scale of all our simulations is 1 min. Hence, the process of cell polarization in our simulations sets in within seconds and is complete within minutes, which is similar to the duration of $\sim$10 min of noise-induced cell polarization in experiments \cite{Maiuri.2015}. Therefore, the noise level in our simulations is enough to destabilize an unpolarized cell in a time scale similar to experimental observations. Thus, the gradient values used in our simulations inform the gradients that should be used in experiments to obtain similarly robust guidance of cell migration against noise.

Moreover, while we studied cell propulsion in a simple fluid, our results can also be interpreted in the context of a cell moving through tissue. 
This broader applicability is supported by our choice of large fluid viscosities (which mimic long-term viscous behavior of tissue) and the observed robustness of our results across various cortical-to-fluid viscosity ratios.

Although simulations were restricted to axisymmetric geometries, the model is fully extensible to general 3D settings. This opens avenues for investigating cell behavior in more complex environments, including branched channels, obstacles, and tissue-like matrices. These situations are relevant for the migration of different cell types, such as immune cells surveying tissues for pathogens or cancer cells invading healthy tissues  \cite{poincloux2011contractility}. Therefore, the guidance mechanisms identified here may help explain how such cells make directional decisions in their complex microenvironments.

Beyond cell migration, our model provides a foundation for studying other biological processes involving active surfaces - for example, cytokinesis, where a contractile ring emerges in the cell cortex and progressively constricts to divide the cell \cite{Satterwhite.1992, Pollard2010}. Minimal models  \cite{Mietke_PRL_2019} suggest that the underlying mechanics are the same as those explored here. 
Having demonstrated that our approach is capable of describing cell-cell interaction, it also opens the door to studying active surface dynamics in multicellular systems, for example, those central to early animal embryogenesis \cite{maitre2016asymmetric,firmin2024mechanics}, an area where computational modeling remains remarkably limited.

Taken together, our results underscore how complex migratory behaviors can emerge from minimal ingredients like cortical contractility, hydrodynamic coupling, and environmental gradients within a unified physical framework. This sets the stage for further theoretical and experimental studies into the principles governing active surface-driven motility.

\section*{Author contributions}
HLG: Formal analysis; Software; Visualization; Writing - original draft; Writing - review \& editing.
CW: Software; Writing - original draft.
PAEH: Writing - review \& editing.
RA: Conceptualization; Supervision; Writing - review \& editing.
SA: Conceptualization; Funding acquisition; Methodology; Supervision; Writing - review \& editing.

\section*{Conflicts of interest}
The authors declare that they have no known competing financial interests or personal relationships that could have appeared to influence the work reported in this paper.

\section*{Data availability}
The numerical simulation code is available at Zenodo (\url{https://doi.org/10.5281/zenodo.17167539}). Installation requires the finite element library AMDiS \cite{amdis2}, which is available online (\url{https://gitlab.math.tu-dresden.de/iwr/amdis}).

\section*{acknowledgments}
SA acknowledges support from DFG (grants 511509575, 328170591, 417223351).
The authors acknowledge computing time on the compute cluster of the Faculty of Mathematics and Computer Science of Technische Universität Bergakademie Freiberg, operated by the computing center (URZ) and funded by the Deutsche Forschungsgemeinschaft (DFG) under DFG grant number 397252409.
This research was supported in part by grant NSF PHY-2309135 and the Gordon and Betty Moore Foundation Grant No. 2919.02 to the Kavli Institute for Theoretical Physics (KITP).

\bibliography{main.bib}

\appendix
\numberwithin{equation}{section}

\section{Supplementary Model Details}

\subsection{Extending a Surface Equation to \texorpdfstring{$\Omega$}{TEXT}} \label{sec:diffuse interface derivation}

In this subsection, we formally justify the diffuse-domain formulation of the surface concentration equation \eqref{eq:phase_field:conc}. 
Therefore, we extend the sharp-interface equation \eqref{eq:c nondim} to the full domain \(\Omega\). To this end, we introduce the weak formulation by testing with a suitable test function \(\psi\) with compact support in \(\Omega \times [0,T]\). This condition implies that \(\psi\) vanishes on the boundary of \(\Omega\) as well as at the initial time \(t=0\) and final time \(t=T\).  
The weak formulation of Eq.~\eqref{eq:c nondim} can then be written as
\begin{align*}
	0 &= \int_\Gamma \psi \left[ \frac{{\rm d}c}{{\rm d}t} + c\nabla_\Gamma \cdot \mathbf{u}- \Delta_\Gamma c \right] \dx\\
	&= \int_\Gamma \frac{{\rm d}}{{\rm d}t} (c\psi) - c\frac{{\rm d}}{{\rm d}t} \psi + c \psi \nabla_\Gamma \cdot {\bf u} +\nabla_\Gamma c\cdot \nabla_\Gamma \psi \dx.
\end{align*}
Using the Reynolds transport theorem on surfaces yields
\begin{align*}
	0 &= \frac{\dd}{\dd t} \int_{\Gamma(t)} \psi c \dx +  \int_{\Gamma} -c\frac{{\rm d}}{{\rm d}t} \psi  +\nabla_\Gamma c\cdot \nabla_\Gamma \psi \dx.
\end{align*}
Now the domain of integration can be extended to $\Omega \supset \Gamma$ by use of the interface distribution $\delta_\Gamma$.  
\begin{align*}
	0 &= \frac{d}{dt} \int_{\Omega} \delta_\Gamma \psi c \dx +  \int_{\Omega} -\delta_\Gamma c\frac{{\rm d}}{{\rm d}t} \psi  + \delta_\Gamma \nabla c \cdot P_\Gamma \cdot \nabla \psi \dx.
\end{align*}
Using the ordinary Reynolds transport theorem (not on a surface) yields
\begin{align*}
	0 &=  \int_{\Omega} \frac{{\rm d}}{{\rm d}t}(\delta_\Gamma c) \psi + \delta_\Gamma c \psi \nabla\cdot {\bf u} + \delta_\Gamma \nabla c\cdot P_\Gamma \cdot \nabla \psi \dx \\
	&=  \int_{\Omega} \partial_t(\delta_\Gamma c) \psi + \nabla\cdot(\delta_\Gamma c {\bf u})\psi - \nabla\cdot(\delta_\Gamma P_\Gamma \nabla c) \psi \dx.
\end{align*}
Going back to the strong formulation gives 
\begin{align*}
	0 &=  \partial_t(\delta_\Gamma c) + \nabla\cdot(\delta_\Gamma c {\bf u}) - \nabla\cdot(\delta_\Gamma P_\Gamma \nabla c) .
\end{align*}
The asymptotic analysis of this equation without the surface projection $P_\Gamma$ shows that the concentration is constant in normal direction, ${\bf n}\cdot \nabla c=0$ for $\epsilon\rightarrow 0$. In this case $P_\Gamma\cdot \nabla c = \nabla c$, hence $P_\Gamma$ can be omitted. 
Approximating the surface delta function by $|\nabla\phi|$ yields the diffuse interface concentration equation \eqref{eq:phase_field:conc}.

\subsection{Asymptotic Analysis} \label{sec:asymptotic}
We will argue in this section, that the solution \(\mathbf{v}(\text{x}, \varepsilon)\) of
\begin{align}
	\vert \nabla \phi \vert \mathbf{v}-\vert \nabla \phi \vert\mathbf{u}-\nabla \cdot [\vert \nabla \phi \vert( \mathbf{ n} \otimes \mathbf{ n}) \nabla \mathbf{v}] &= \mathbf{0} &\text{on }\Omega,\tag{\ref{eq:phase_field:v_extended}}
\end{align}
is a constant normal extension of ${\bf u}$, i.e. for \(\varepsilon\to 0\), the equations converge formally to
\begin{align}
	\mathbf{v} &= \mathbf{u}&\text{on }\Gamma,\tag{\ref{eq:sharp_interface:u=v}}\\
	(\mathbf{n}\cdot\nabla) \mathbf{v}  &= 0 &\text{on }\Gamma\tag{\ref{eq:sharp_interphase:v'=0}}.
\end{align}
To do so, we utilize matched asymptotic analysis and  follow the argumentation of Ref.\cite{RaetzVoigt_2006} and Ref.\cite{li2009_diffuse_domain}. We start by introducing a new inner coordinate system in a neighborhood of the surface \(\Gamma\), such that for any \(\rm x\) in the neighborhood there is an unique \(s(\text{x})\in\Gamma\) with minimal distance to \( \rm x\). Then \( \rm x\) can be represented as
\begin{displaymath}
	\text{x} = s( \text{x}) + r( \text{x})\mathbf{n} = s(\text{x}) + \varepsilon z(\text{x}) \mathbf{n},
\end{displaymath}
where \(r\) is a signed distance function with \(r<0\) in \(\Omega_1\) and \(r>0\) in \(\Omega_0\). The variable $z$ is a scaled distance function defined by \(z(\text{x})=\frac{r(\text{x})}{\varepsilon}\).
The phase-field function can be expressed in terms of these coordinates as 
\begin{align}
	\phi
	= \frac{1}{2} \left(1-\tanh\left(\frac{r}{\sqrt{2}\varepsilon}\right)\right) 
	= \frac{1}{2} \left(1-\tanh\left(\frac{z}{\sqrt{2}}\right)\right). \label{eq:phi of z}
\end{align}

We expand \(\mathbf{u}(\text{x}, \varepsilon)\) and \(\mathbf{v}(\text{x}, \varepsilon)\) outside the interface region in terms of the original coordinate system \(\mathbf{u} = \sum_{i=0}^\infty \varepsilon^i u^{(i)}, \mathbf{v} = \sum_{i=0}^\infty \varepsilon^i v^{(i)}\).  
This procedure is referred to as the outer expansion.
In the neighborhood of \(\Gamma\) we introduce for \(\mathbf{\hat u}(s, z)\) and \(\mathbf{\hat v}(s,  z)\)  the inner expansion \(\mathbf{\hat u} = \sum_{i=0}^\infty \varepsilon^i \hat u^{(i)}, \mathbf{\hat v} = \sum_{i=0}^\infty \varepsilon^i \hat v^{(i)}\) in terms of the new coordinate system. In an overlapping region, both are valid representations of the same function and thus the following matching conditions hold
\begin{align}
	& \lim_{r\to \pm 0} u^{(0)}(s,r) &=& \lim_{z\to\pm\infty} \hat u^{(0)}(s,z)\label{eq:A:matching4}\\
	& \lim_{r\to \pm 0} v^{(0)}(s,r) &=& \lim_{z\to\pm\infty} \hat v^{(0)}(s,z)\label{eq:A:matching5}\\
	& \lim_{r\to \pm 0} \nabla v^{(0)}(s,r) \cdot \mathbf{n} &=& \lim_{z\to\pm\infty} \partial_z \hat v^{(1)}(s,z).\label{eq:A:matching6}
\end{align}
Inserting the \textbf{outer expansions} into Eq.\eqref{eq:phase_field:v_extended} then yields
\begin{align*}
	0=\vert \nabla \phi^{(0)} \vert v^{(0)}-\vert \nabla \phi^{(0)} \vert u^{(0)}-\nabla \cdot [\vert \nabla \phi^{(0)} \vert( \mathbf{n} \otimes \mathbf{n}) \nabla v^{(0)}],
\end{align*}
which gives the trivial identity $0=0$ away from the interface. 

In the inner coordinate system, Eq.\eqref{eq:phase_field:v_extended} turns to
\begin{align*}
    0 &= \vert \nabla \phi \vert\, \mathbf{\hat v} 
         - \vert \nabla \phi \vert\, \mathbf{\hat u} \\
      &\quad - \left( \frac{1}{\varepsilon} \mathbf{n} \partial_z + \nabla_\Gamma \right) \cdot 
         \left[ \vert \nabla \phi \vert\, (\mathbf{n} \otimes \mathbf{n}) 
         \left( \frac{1}{\varepsilon} \mathbf{n} \partial_z + \nabla_\Gamma \right) \mathbf{\hat v} \right] \\
      &= \vert \nabla \phi \vert\, \mathbf{\hat u} 
         - \vert \nabla \phi \vert\, \mathbf{\hat v} \\
      &\quad + \frac{1}{\varepsilon^2} \partial_z \left( \vert \nabla \phi \vert\, \partial_z \mathbf{\hat v} \right) 
         + \frac{1}{\varepsilon} \nabla_\Gamma \cdot \left( \vert \nabla \phi \vert\, \mathbf{n} \partial_z \mathbf{\hat v} \right),
\end{align*}
where we used ${\bf n}\cdot\nabla_\Gamma=0$ and $\partial_z {\bf n}=0$. 
By inserting the \textbf{inner expansion} and comparing powers of \(\varepsilon\), we obtain the following condition at order $1/\varepsilon^2$
\begin{align*}
	0= \partial_z \left(\vert \nabla \phi\vert \partial_z \hat v^{(0)}\right).
\end{align*}
Thus, \(\vert \nabla \phi\vert \partial_z \hat v^{(0)}\) is constant in $z$. This constant value must be zero since $|\nabla\phi|\rightarrow 0$ for $z\rightarrow\pm\infty$ (see Eq.\eqref{eq:phi of z}). 
Hence, \(\partial_z {\hat v^{(0)}}=0\). Using this result, we obtain at order $1/\varepsilon$
\begin{align*}
	0 = \partial_z \left(\vert \nabla \phi\vert \partial_z \hat v^{(1)}\right).
\end{align*}
Similarly to above, we deduce \(\partial_z \hat v^{(1)}=0\). Using matching condition \eqref{eq:A:matching6} we obtain the desired condition \eqref{eq:sharp_interphase:v'=0}, as
\begin{align*}
	(\mathbf{n}\cdot\nabla) \mathbf{v}^{(0)}  &= 0 & {\rm on~}\Gamma.
\end{align*}
Finally, we have at order $1$ 
\begin{align*}
	0=&\vert \nabla \phi \vert \hat u^{(0)}-\vert \nabla \phi \vert \hat v^{(0)}+\partial_z \left(\vert \nabla \phi\vert \partial_z \hat v^{(2)}\right) \\
    &\quad + \nabla_\Gamma \cdot \left(\vert \nabla \phi\vert \mathbf{n} \partial_z \hat v^{(1)}\right)\\
	=&\vert \nabla \phi \vert \hat u^{(0)}-\vert \nabla \phi \vert \hat v^{(0)}+\partial_z \left(\vert \nabla \phi\vert \partial_z \hat v^{(2)}\right).
\end{align*}
Integrating from \(z=-\infty\) to \(z=+\infty\) yields
\begin{align*}
	0 =& \int_{-\infty}^{+\infty} \left\{\vert \nabla \phi \vert \hat u^{(0)}-\vert \nabla \phi \vert \hat v^{(0)}+\partial_z \left(\vert \nabla \phi\vert \partial_z \hat v^{(2)}\right) \right\} \dd z\\
	=& ~\hat u^{(0)}\int_{-\infty}^{+\infty} \vert \nabla \phi \vert \dd z  - \hat v^{(0)} \int_{-\infty}^{+\infty} \vert \nabla \phi \vert \dd z \\
    &\quad + \left[ \vert \nabla \phi\vert \partial_z \hat v^{(2)}\right]_{-\infty}^{+\infty},
\end{align*} 
where we used that \(\hat v^{(0)}\) is constant in $z$ as shown above and \(\hat u^{(0)}\) is constant in $z$ since ${\bf u}$ is a continuous function (in fact: $\hat u^{(0)} = {\bf u}_{|\Gamma}$). Further, from Eq.\eqref{eq:phi of z} we conclude for $z=\pm\infty$ that  $|\nabla\phi|=0$, hence \(\left[ \vert \nabla \phi\vert \partial_z \hat v^{(2)}\right]_{-\infty}^{+\infty}=0\). 
Dividing by the (non-zero) integrals, we obtain $\hat v^0= \hat u^0$. With conditions \eqref{eq:A:matching4}-\eqref{eq:A:matching5} we conclude $v^0=  u^0$, from which we recover the desired Eq.\eqref{eq:sharp_interface:u=v}.

\subsection{Time Discretization} \label{sec:time}

After establishing the equation system, it remains to solve the six strongly coupled equations. 
To reduce the size of the linear equation system that we have to solve in each time step, we decouple them and solve the Stokes-Cahn-Hilliard equations independently from the velocity extension equation and the concentration equation.

We employ the stable linear semi-implicit time discretization from Ref.\cite{Aland_2014_time} for solving the coupled Stokes-Cahn-Hilliard-Navier system in the n-th time step

\begin{widetext}
\begin{align*}
	\frac{\phi^n-\phi^{n-1}}{\Delta t} =& \nabla \cdot (M \nabla \mu^n) - \nabla \phi^n \cdot \mathbf{v}^{n-1},\\
	\mu^n =& -\varepsilon \Delta \phi^n + \varepsilon^{-1}[W'(\phi^{n-1}) + W''(\phi^{n-1})(\phi^n-\phi^{n-1})] ,\\
	-\nabla \cdot \left[\eta(\phi^{n-1})(\nabla \mathbf{u}^n + (\nabla \mathbf{u}^n)^T )\right] + \nabla p^n =&
	\Pe\left(3\sqrt{2}\sigma(c^{n-1})\mu^n\nabla\phi^{n-1} + |\nabla\phi^{n-1}| \sigma'(c^{n-1}){\tilde \nabla}^{n-1}_\Gamma c^{n-1}\right) \\
	+ \nabla \cdot \left[|\nabla\phi^{n-1}|(1-\nu){\tilde P}^{n-1}_\Gamma{\tilde \nabla}^{n-1}_\Gamma\right.&\cdot \mathbf{u}^n +\left.\nu |\nabla\phi^{n-1}| {\tilde P}^{n-1}_\Gamma(\nabla \mathbf{u}^n + \nabla (\mathbf{u}^n)^T){\tilde P}^{n-1}_\Gamma  \right],\\
	\nabla\cdot\mathbf{u}^n=& 0. 
\end{align*}
where we calculated \(\phi^{n-1}\), \(\mathbf{v}^{n-1}\) and \(c^{n-1}\) in the previous time step. 

Using the current time steps \(\phi^n\) and \(\mathbf{u}^n\), we obtain the projected velocity \(\mathbf{v}\) from
\begin{align*}
	\vert \nabla \phi^{n} \vert \mathbf{v}^n -\nabla \cdot \left(\vert \nabla \phi^{n} \vert({ \mathbf{\tilde n}}^n \otimes { \mathbf{\tilde n}}^n) \nabla \mathbf{v}^n \right) = \vert \nabla \phi^{n} \vert\mathbf{u}^{n}.
\end{align*}

Finally, we solve the concentration equation using the previously computed values for \(\phi^n\) and \(\mathbf{v}^n\)
\begin{align*}
	\frac{ \vert \nabla \phi^{n}\vert c^n - \vert \nabla \phi^{n-1}\vert c^{n-1}}{\Delta t}+ \nabla \cdot (\vert \nabla \phi^n \vert c^n \mathbf{v}^{n})- \nabla \cdot (\vert \nabla \phi^{n}\vert \nabla c^n) &=\alpha \vert \nabla \phi^n \vert (m(t^{n-1})-m(0)).
\end{align*}
Note that, even without the term on the right hand side (i.e. in case $\alpha=0$), the proposed time discretization ensures mass conservation on the discrete level. This property becomes obvious in the weak from of the equation with a test function \(\varphi \in H^1(\Omega)\):
\begin{align*}
	\int_\Omega \varphi \vert \nabla \phi^{n}\vert c^n \dx- \Delta t \int_\Omega \nabla \varphi \cdot \left[(\vert \nabla \phi^n \vert c^n \mathbf{v}^{n})- (\vert \nabla \phi^{n}\vert \nabla c^n)\right] \dx = \int_\Omega \varphi \vert \nabla \phi^{n-1}\vert c^{n-1}\dx.
\end{align*}
\end{widetext}
This holds especially for \(\varphi\equiv 1\) and thus
\begin{align*}
	\int_\Omega \vert \nabla \phi^{n}\vert c^n \dx = \int_\Omega \vert \nabla \phi^{n-1}\vert c^{n-1} \dx
\end{align*}
which is the phase-field equivalent of exact mass conservation 
\begin{align*}
	\int_{\Gamma^n} c^n \dx = \int_{\Gamma^{n-1}} c^{n-1} \dx.
\end{align*}
However, due to adaptive grid refinement and coarsening, small errors in surface mass may accumulate over time, requiring the mass correction ($\alpha>0$) for long simulation times. 

\subsection{Space Discretization} \label{sec:space}

We solve the system of equations in each time step with a Finite Element method based on the Finite Element toolboxes DUNE  \cite{Sander_2020} and AMDiS  \cite{amdis2,VeyVoigt_2006,Witkowski_2015}. To decrease the size of the system, we avoid solving the full 3D-problem and assume that the cell shape and myosin concentration distribution are  axisymmetric. This holds in particular for the biologically most relevant patterns, which are rings and single spots   of increased concentration. The assumption of axisymmetry reduces computations effectively to a 2D domain from which the full 3D-solution can be recovered by rotating the calculated 2D solution 360 degrees. Detailed explanations can be found in the Appendix \ref{app:axisymmetric}.

An adaptive grid is employed to accurately resolve the phase field and surface forces, see Fig. \ref{fig:sketch} (bottom right). Interfacial grid refinement is heuristically chosen, based on the value of the phase field, such that the grid size is \( h_{\text{int}}\) where \(0.05<\phi<0.95\) and \( h_{\text{bulk}}\) otherwise. 

The numerical approximations \(\mathbf{\tilde n}\) and \(\tilde P_\Gamma\) become less accurate when evaluated farther from the interface. To avoid numerical errors accumulating in the outer areas, we replace \(\vert \nabla \phi \vert\) by
\[\vert \nabla \phi \vert^* \coloneqq \begin{cases} \vert \nabla \phi \vert, & \text{if } \vert \nabla \phi \vert>10^{-4}\\ 0, &\text{otherwise.}\end{cases}\]
Interchanging \(\vert \nabla \phi \vert^*\) with \(\max(\vert \nabla \phi \vert^*, 10^{-1})\) in the diffusion terms in equations \eqref{eq:phase_field:conc} and \eqref{eq:phase_field:v_extended}, we ensure that the induced linear systems remain regular.
For the adjusted problem and the introduced discretization, we use Lagrange-P2 elements for the phase field \(\phi\), the chemical potential \(\mu\), the velocities \(\mathbf{u}\) and \(\mathbf{v}\) and the concentration \(c\), only for the pressure \(p\) we use a P1-ansatz space.

\subsection{Axisymmetric Formulation} \label{app:axisymmetric}

As hinted before we use the rotationally symmetric setup to our advantage and save computation time by solving a 2D-problem and then rotate the solution around the \(z\)-axis turning it into 3D. For this we consider cylinder coordinates \((\rho, \varphi, z)\) throughout this section. A vector \(\boldsymbol{u} \in \mathbb{R}^3\) is then represented by \(\boldsymbol{u} = u_\rho \boldsymbol{e}_\rho+ u_\varphi \boldsymbol{e}_\varphi + u_z \boldsymbol{e}_z\) and we write \(\boldsymbol{u}=(u_\rho, u_\varphi, u_z)^T\) and use that notation for all vectors and matrices following.

To reduce the complexity of the problem, we utilize that the solution will be axisymmetric, and thus constant in \(\varphi\)-direction. We therefore replace the usual 3D-Cartesian differential operators by their cylindrical equivalent without the \(\varphi\)-direction.

We denote the unit vectors in cylindrical coordinates by \(\boldsymbol{e}_\rho\), \(\boldsymbol{e}_\varphi\) and \(\boldsymbol{e}_z\). It holds
\begin{align*}
	&\partial_\rho \boldsymbol{e}_\rho = 0, && \partial_\varphi \boldsymbol{e}_\rho = \boldsymbol{e}_\varphi, &&\partial_z \boldsymbol{e}_\rho = 0,\\
	&\partial_\rho \boldsymbol{e}_\varphi = 0, && \partial_\varphi \boldsymbol{e}_\varphi = -\boldsymbol{e}_\rho, &&\partial_z \boldsymbol{e}_\varphi = 0,\\
	&\partial_\rho \boldsymbol{e}_z = 0, && \partial_\varphi \boldsymbol{e}_z = 0, &&\partial_z \boldsymbol{e}_z = 0.
\end{align*}

In cylindrical coordinates, the gradient is given by 
\begin{displaymath}
	\nabla_R \coloneqq \left[\boldsymbol{e}_\rho \partial_\rho + \boldsymbol{e}_\varphi(\frac{1}{\rho} \partial_\varphi)  + \boldsymbol{e}_z \partial_z \right] \otimes.
\end{displaymath}

Specifically, the gradient of a scalar field \(f\) is 
\begin{widetext}
\begin{displaymath}
	\nabla_R f = \partial_\rho f \boldsymbol{e}_\rho + (\frac{1}{\rho} \partial_\varphi f) \boldsymbol{e}_\varphi + \partial_z f \boldsymbol{e}_z = (\partial_\rho f, \frac{1}{\rho}\partial_\varphi f, \partial_z f)^T
\end{displaymath}
and the gradient of a vector field \(\boldsymbol u=u_\rho \boldsymbol{e}_\rho+ u_\varphi \boldsymbol{e}_\varphi + u_z \boldsymbol{e}_z\)

\begin{displaymath}
	\nabla_R \boldsymbol u =\boldsymbol{e}_\rho  \otimes(\partial_\rho \boldsymbol u )  + \boldsymbol{e}_\varphi \otimes (\frac{1}{\rho} \partial_\varphi \boldsymbol u) + \boldsymbol{e}_z \otimes (\partial_z \boldsymbol u)  =
	\begin{pmatrix}
		\partial_\rho u_\rho & \partial_\rho u_\varphi & \partial_\rho u_z \\
		\frac{1}{\rho}(\partial_\varphi u_\rho - u_\varphi) & \frac{1}{\rho}(\partial_\varphi u_\varphi + u_\rho) & \frac{1}{\rho} \partial_\varphi u_z  \\
		\partial_z u_\rho & \partial_z u_\varphi & \partial_z u_z
	\end{pmatrix}.
\end{displaymath}
The divergence is defined in a similar fashion as the inner product
\begin{displaymath}
	\nabla_R \cdot \coloneqq \left[\boldsymbol{e}_\rho \partial_\rho + \boldsymbol{e}_\varphi (\frac{1}{\rho} \partial_\varphi) + \boldsymbol{e}_z \partial_z \right] \cdot,
\end{displaymath}
thus the divergence of a vector field \(\boldsymbol u\) is given by
\begin{displaymath}
	\nabla_R \cdot \boldsymbol{u} = \partial_\rho u_\rho + \frac{1}{\rho} (\partial_\varphi u_\varphi + u_\rho) + \partial_z u_z. 
\end{displaymath}
We observe for a tensor \(A\in\mathbb{R}^{n\times n}\)
\begin{align*}
	\nabla_R \cdot A &= \nabla_R \cdot 
	\begin{pmatrix}
		a_{\rho,\rho} & a_{\rho,\varphi} & a_{\rho,z}\\
		a_{\varphi,\rho} & a_{\varphi,\varphi} &  a_{\varphi,z}\\
		a_{z,\rho} & a_{z,\varphi} & a_{z,z}
	\end{pmatrix}\\
	&= \sum_{k,l=\rho, \varphi, z} \boldsymbol{e}_\rho \cdot \partial_\rho (a_{k,l} \boldsymbol{e}_k\otimes \boldsymbol{e}_l) 
	+ \sum_{k,l=\rho, \varphi, z} \boldsymbol{e}_\varphi \cdot \frac{1}{\rho} \partial_\varphi (a_{k,l} \boldsymbol{e}_k\otimes \boldsymbol{e}_l) 
	+ \sum_{k,l=\rho, \varphi, z} \boldsymbol{e}_z \cdot \partial_z (a_{k,l} \boldsymbol{e}_k\otimes \boldsymbol{e}_l)\\
	&= \sum_{l=\rho, \varphi, z} (\partial_\rho a_{\rho,l}) \boldsymbol{e}_l 
	+ \frac{1}{\rho} \sum_{l=\rho, \varphi, z} (\partial_\varphi a_{\varphi,l} + a_{\rho,l}) \boldsymbol{e}_l +a_{\varphi,l} (\partial_\varphi \boldsymbol{e}_l)
	+ \sum_{l=\rho, \varphi, z} (\partial_z a_{z,l}) \boldsymbol{e}_l\\
	&= \sum_{\substack{k=\rho, z \\ l=\rho, \varphi, z}} (\partial_k a_{k,l}) \boldsymbol{e}_l 
	+ \frac{1}{\rho} \sum_{l=\rho, \varphi, z} (\partial_\varphi a_{\varphi,l} + a_{\rho,l}) \boldsymbol{e}_l + \frac{1}{\rho} a_{\varphi,\rho} \boldsymbol{e}_\varphi -\frac{1}{\rho} a_{\varphi, \varphi} \boldsymbol{e}_\rho.
\end{align*}
Furthermore, the normal vector on the surface \(\Gamma\) is of the form \(\boldsymbol n = n_\rho \boldsymbol{e}_\rho+n_z\boldsymbol{e}_z\) and 
\begin{displaymath}
	P_\Gamma = \begin{pmatrix}1-n_\rho n_\rho & 0& -n_\rho n_z\\0&1&0\\-n_\rho n_z&0&1-n_zn_z \end{pmatrix} 
	= \begin{pmatrix}p_{\rho,\rho} & 0& p_{\rho,z}\\0&p_{\varphi,\varphi}&0\\p_{\rho,z}&0&p_{z,z} \end{pmatrix} .
\end{displaymath}
Now we utilize that our system is constant in azimuthal direction, in particular all first derivatives in \(\varphi\)-direction vanish and \(u_\varphi=0\). With that, the incompressibility condition \eqref{eq:NS 2 nondim} turns to 
\begin{displaymath}
	\nabla_R\cdot \boldsymbol{u} = \left[\partial_\rho \boldsymbol{e}_\rho + (\frac{1}{\rho} \partial_\varphi) \boldsymbol{e}_\varphi + \partial_z \boldsymbol{e}_z\right] \cdot \boldsymbol u = \partial_\rho u_\rho + \frac{1}{\rho}\left(\partial_\varphi u_\varphi + u_\rho\right) +\partial_z u_z \overset{u_\varphi=0}{=} \frac{1}{\rho}\partial_\rho\left(\rho u_\rho \right) + \partial_z u_z.
\end{displaymath}
For the conservation of momentum, Eq.\eqref{eq:NS 1 phase-field},   
\begin{align}
	-\nabla \cdot \left[\eta(\nabla \mathbf{u} + \nabla \mathbf{u}^T )\right] + \nabla p =&
	\nabla \cdot \left[\delta_\Gamma(1-\nu)P_\Gamma{\tilde\nabla}_\Gamma \cdot \mathbf{u} +\nu \delta_\Gamma P_\Gamma(\nabla \mathbf{u} + \nabla \mathbf{u}^T)P_\Gamma  \right] \nonumber\\
	&+\Pe\delta_\Gamma \left(3\sqrt{2} \sigma(c) \mu\nabla\phi + P_\Gamma { \nabla} \sigma(c)\right).
\end{align}
Note that for shorter notation we write here $\delta_\Gamma$ instead of the $|\nabla\phi|$. 
Now formulating this in the described axisymmetric setting, yields for the first order derivatives
\begin{align*}
	\nabla_R p &= (\partial_\rho p) \boldsymbol{e}_\rho + (\partial_z p) \boldsymbol{e}_z\\
	\nabla_R \phi &= (\partial_\rho \phi) \boldsymbol{e}_\rho + (\partial_z \phi) \boldsymbol{e}_z\\
	P_\Gamma \nabla\sigma(c) &= \sigma'(c) P_\Gamma \nabla_R c = \sigma'(c) P_\Gamma ((\partial_\rho c) \boldsymbol{e}_\rho + (\partial_z c) \boldsymbol{e}_z)
\end{align*}
and for the second order derivatives
\begin{align*}
	&\nabla_R \cdot \left[\eta(\nabla_R \mathbf{u} + \nabla_R \mathbf{u}^T )\right] 
	&=& \nabla_R \cdot \eta
	\begin{pmatrix}
		2\partial_\rho u_\rho & 0 & \partial_z u_\rho+\partial_\rho u_z\\
		0 & \frac{2}{\rho} u_\rho & 0 \\
		\partial_\rho u_z+\partial_z u_\rho & 0 & 2\partial_z u_z
	\end{pmatrix}\\
	&&=& \sum_{k,l=\rho, z} \partial_k(\eta \partial_l u_k+\eta\partial_k u_l) \boldsymbol{e}_l 
	+ \frac{\eta}{\rho} \sum_{l=\rho, z} (\partial_l u_\rho+\partial_\rho u_l) \boldsymbol{e}_l -\frac{2\eta}{\rho^2} u_\rho \boldsymbol{e}_\rho,
	\\
	&\nabla_R \cdot \left[\nu\delta_\Gamma P_\Gamma(\nabla_R \mathbf{u} + \nabla_R \mathbf{u}^T )P_\Gamma\right] 
	&=& \nabla_R \cdot \nu \delta_\Gamma
	\left( \sum_{\substack{i,j=\rho, z\\k,l=\rho, z}} p_{k,i}(\partial_i u_j + \partial_j u_i)p_{j,l} \boldsymbol{e}_k\otimes \boldsymbol{e}_l + p_{\varphi,\varphi}\frac{2}{\rho} u_\rho p_{\varphi,\varphi} \boldsymbol{e}_\varphi\otimes \boldsymbol{e}_\varphi\right)\\
	&&=& \sum_{k,l=\rho, z} \partial_k
	\left(\nu \delta_\Gamma \sum_{i,j=\rho, z} p_{k,i}(\partial_i u_j + \partial_j u_i)p_{j,l}\right) \boldsymbol{e}_l \\
	&&&+ \frac{\nu \delta_\Gamma}{\rho} \sum_{l=\rho, z} \left(\sum_{i,j=\rho, z} p_{\rho,i}(\partial_i u_j + \partial_j u_i)p_{j,l}\right) \boldsymbol{e}_l -\frac{2\nu\delta_\Gamma}{\rho^2} u_\rho \boldsymbol{e}_\rho,
	\\
	&\nabla_R\cdot\left[(1-\nu)\delta_\Gamma {\tilde\nabla}_{\Gamma,R} \cdot \boldsymbol{u} P_\Gamma\right]
	&=&\nabla_R\cdot (1-\nu)\delta_\Gamma \left(\sum_{i,j=\rho,z} p_{i,j}\partial_i u_j + p_{\varphi,\varphi}\frac{1}{\rho}u_\rho\right) P_\Gamma\\
	&&=&\sum_{k,l=\rho, z} \partial_k
	\left((1-\nu)\delta_\Gamma \left(\sum_{i,j=\rho,z} p_{i,j}\partial_i u_j + \frac{u_\rho}{\rho}\right) p_{k,l}\right) \boldsymbol{e}_l \\
	&&&+ \frac{(1-\nu) \delta_\Gamma}{\rho} \sum_{l=\rho, z} \left(\sum_{i,j=\rho,z} p_{i,j}\partial_i u_j  + \frac{u_\rho}{\rho} \right)p_{\rho,l} \boldsymbol{e}_l \\
	&&&-\frac{(1-\nu)\delta_\Gamma}{\rho} \left(\sum_{i,j=\rho,z} p_{i,j}\partial_i u_j  + \frac{u_\rho}{\rho} \right) \boldsymbol{e}_\rho.
\end{align*}
\end{widetext}

\begin{figure*}
    \centering
    \includegraphics[width=1\textwidth]{figures/epsilon.pdf}
    \caption{\textbf{- Influence of interface thickness $\varepsilon$ on cell adaptation and cell speed.}     
\textbf{Left:} Cell contour at $t=10.0$ for three interface thickness values ($\varepsilon=0.04,0.02,0.01$), showing that the thickness of the thin fluid film between cell and wall is proportional to $\varepsilon$. 
\textbf{Right:} Cell speed $u_{\text{cell}}$ over time for different interface thicknesses $\varepsilon$. Smaller $\varepsilon$ values lead to higher steady-state speeds, while the results converge for sufficiently small $\varepsilon$.    }
    \label{fig:epsilon}
\end{figure*}

Finally, for the concentration equation \eqref{eq:phase_field:conc} the application of rotational operators $\nabla_R$ and $\nabla_R\cdot$ is straightforward. 
Similarly, for the Cahn-Hilliard equation, $\Delta \phi$ is replaced by $\nabla_R\cdot \nabla_R \phi$, which is necessary for the chemical potential to approximate the correct 3-dimensional curvature. However, we refrain from using axisymmetric operators for $\Delta\mu$ to better preserve cell volume. Imposing rotational symmetry in this term would effectively increase the extracellular-to-intracellular volume ratio, thereby promoting diffusion of the smaller intracellular phase into the larger extracellular domain - a known artifact in phase-field models  \cite{yue2007spontaneous}. Since our objective is to conserve intracellular volume, such rotational symmetry is intentionally avoided in this specific term, thereby lowering droplet loss of mass by decreasing the volume contrast of intracellular to extracellular domains, due to two-dimensional instead of three-dimensional domains.
Note that since the primary purpose of the phase field is to track the cellular interface, rotationally symmetric operators are not necessary for $\Delta \mu$; the equations still approximate the same sharp interface limit.

\subsection{Influence of the Interface Thickness \texorpdfstring{$\varepsilon$}{TEXT}}\label{sec:epsilon}

To evaluate the effect of the interface thickness parameter $\varepsilon$, we performed two validation studies.
In the first study, we compared the initial cell adaptation to the channel for three different values of $\varepsilon$. 
The simulation results (Fig.~\ref{fig:epsilon} (left)) indicate that the thickness of the fluid film separating the cell from the wall increases with the interface thickness parameter $\varepsilon$, consistent with the expected scaling behavior of phase-field models.

The second validation study investigates the influence of $\varepsilon$ on the velocity of a cell that has already been initialized with a  prescribed shallow 1-mode deformation, thereby avoiding additional effects from pattern formation. The comparison across different values of $\varepsilon$ shows that the cell velocity converges as $\varepsilon$ decreases, approaching a constant nonzero value. As expected, cells move slightly faster for smaller $\varepsilon$, since a thinner fluid layer between the cell and the wall results in increased shear interaction and, consequently, a higher net driving force.

Overall, these studies confirm that, for sufficiently small $\varepsilon$, the simulation results are largely independent of the precise choice of the interface thickness, corroborating the robustness of the findings presented in the main text.

\begin{figure*}
    \centering
    \includegraphics[width=1\textwidth]{figures/peclet.pdf}
    \caption{\textbf{- Emergence of polar 1-mode concentration patterns depends on the {P\'eclet} number.}  
    \textbf{Left:} Centered concentration profile along the axial coordinate $z$ for different {P\'eclet} numbers at $t = 10$. For $\Pe < 20$, profiles remain nearly left-right symmetric, while for $Pe \geq 20$, distinct polar patterns emerge. The slight visible 2-mode pattern results from the presence of the channel walls and the corresponding initial adaption of cell shape. 
    \textbf{Right:} Time evolution of the left-right concentration difference, $c_R - c_L$, for varying {P\'eclet} numbers. Above the critical value ($\text{Pe} \geq 20$), a persistent symmetry breaking and increasing polarity between cell ends is observed, while lower {P\'eclet} numbers result in negligible polarization.}
    \label{fig:peclet}
\end{figure*}

\subsection{Critical {P\'eclet} Number and Pattern Formation} \label{sec:peclet}

To assess the onset of spontaneous polarization in our model, we performed a systematic set of simulations for different {P\'eclet} numbers $\Pe$. Each simulation was initialized with a slight 1-mode perturbation (amplitude 0.002) and we monitor both the spatial concentration profile along the axial coordinate and the concentration difference between the two cell ends over time.

Figure~\ref{fig:peclet} (left) shows that for {P\'eclet} numbers below 15, the concentration profiles remain nearly left-right symmetric. Above $\Pe \approx 20$, polar patterns consistently emerge, characterized by high concentration at one cell end and low concentration at the other. 
The slight visible 2-mode pattern results from the presence of the channel walls and the corresponding initial adaption of cell shape. This 2-mode, however, does not further increase over time, indicating that the critical $\Pe$ number is larger than 25 for this mode. Higher order modes are not observed to grow in all considered cases. 

Figure~\ref{fig:peclet} (right) quantifies the evolution of the concentration difference between the right and left ends of the cell, $c_{R}-c_{L}$. For $\Pe < 20$, 1-mode polarity remains negligible, while for $\Pe \ge 20$, a pronounced and persistent increase is observed, signifying spontaneous emergence of polar 1-mode patterns.
The establishment of a slight concentration difference for $\Pe=20$ indicates that this value is very close to the critical {P\'eclet} number of the 1-mode.

\end{document}